\PassOptionsToPackage{pdfpagelabels=false}{hyperref}
\documentclass[useAMS,usenatbib]{mnras}
\pdfoutput=1
\usepackage[pdftex]{graphicx}
\usepackage{url}
\usepackage{amsmath}
\usepackage{natbib}
\usepackage{geometry}
\usepackage{txfonts}
\usepackage{times}

\def \MSUN{{\rm M}_{\odot}}

\title[Azimuthal Anisotropy in CGM Magnetic Fields]{Azimuthal Anisotropy of Magnetic Fields in the Circumgalactic Medium Driven by Galactic Feedback Processes}
\author[R. Ramesh et al.]{Rahul Ramesh$^{1}$\thanks{E-mail: rahul.ramesh@stud.uni-heidelberg.de}, Dylan Nelson$^{1}$, Volker Heesen$^{2}$ and Marcus Brüggen$^{2}$
\\
$^{1}$ Universität Heidelberg, Zentrum für Astronomie, Institut für theoretische Astrophysik, Albert-Ueberle-Str. 2, 69120 Heidelberg, Germany\\
$^{2}$ Hamburg University, Hamburger Sternwarte, Gojenbergsweg 112, 21029 Hamburg, Germany\\
}

\date{}
\pubyear{2023}

\begin{document}

\maketitle

\begin{abstract}
We use the TNG50 cosmological magnetohydrodynamical simulation of the IllustrisTNG project to show that magnetic fields in the circumgalactic medium (CGM) have significant angular structure. This azimuthal anisotropy at fixed distance is driven by galactic feedback processes that launch strong outflows into the halo, preferentially along the minor axes of galaxies. These feedback-driven outflows entrain strong magnetic fields from the interstellar medium, dragging fields originally amplified by small-scale dynamos into the CGM. At the virial radius, $z=0$ galaxies with M$_\star$\,$\sim$\,$10^{10}\,\rm{M_\odot}$ show the strongest anisotropy ($\sim 0.35$~dex). This signal weakens with decreasing impact parameter, and is also present but weaker for lower mass as well as higher mass galaxies. Creating mock Faraday rotation measure (RM) sightlines through the simulated volume, we find that the angular RM trend is qualitatively consistent with recent observational measurements. We show that rich structure is present in the circumgalactic magnetic fields of galaxies. However, TNG50 predicts small RM amplitudes in the CGM that make detection difficult as a result of other contributions along the line of sight.
\end{abstract}


\section{Introduction}\label{intro}

Observational and theoretical studies suggest that galaxies are surrounded by a halo of gas that typically extends out to roughly the virial radius of their parent dark matter halos. Termed the circumgalactic medium (CGM), this multi-scale, multi-phase reservoir of gas is believed to play a critical role in the evolution of galaxies (see \citealt{donahue2022} for a recent review of the CGM).

Simultaneously, the CGM is affected by physical processes that take place within the galaxy. For instance, outflows driven by the central supermassive black hole (SMBH) can launch gas out of high-mass galaxies with $M_\star \gtrsim 10^{10.5}\,\MSUN$ and into the CGM at high velocities \citep{oppenheimer2020, ramesh2023a}, increasing cooling times and preventing future accretion and star formation \citep{zinger2020, davies2020}. They can create bubbles of hot, rarified gas \citep{pillepich2021} similar to the Fermi/eROSITA bubbles emerging from the galactic centre of the Milky Way \citep{su2010, predehl2020}. For less massive galaxies, outflows driven by supernovae and stellar winds dominate \citep{fielding2017, li2020}, reshaping the CGM with significant inputs of mass, momentum, energy, and metals \citep{nelson2019, mitchell2020, pandya2021}.

Cosmological galaxy formation simulations tend to find that outflows are anisotropic, preferentially propagating perpendicular to the galactic disk, i.e along directions where the density of gas is lower than in the disk \citep[e.g][]{nelson2019}. As a result, the CGM is predicted to possess angular anisotropies in key physical quantities including metallicity \citep{peroux2020}, density, and temperature \citep{truong2021}. The star formation activity of satellite galaxies is expected to respond to this gas anisotropy with an `angular conformity' signal \citep{martinnavarro2021}. This picture of an azimuthally anisotropic CGM has also received observational support with inferences of angular dependencies in metallicity \citep{cameron2021} and density \citep{zhang2022} profiles.

In addition to hydrodynamical and thermal gas processes, non-thermal components including magnetic fields may play a key role in the CGM. Theoretical studies suggest that they can lengthen the lifetimes of small, cold gas clouds by suppressing fluid instabilities \citep{ji2018,berlok2019,sparre2020}, or by contributing support in the form of magnetic pressure \citep{nelson2020, ramesh2023b}. It has recently become possible to probe the impact of magnetic fields in large-volume simulations of cosmic structure formation that self-consistently include magnetohydrodynamics within the context of a realistic galaxy population -- IllustrisTNG is a notable example. Simulations have shown how magnetic fields may influence the propreties of halo gas \citep{pakmor2020,vandevoort2021}, outflows \citep{steinwandel2020}, and the hot intracluster medium \citep{vazza2014}.

Observationally, extragalactic magnetic fields are difficult to measure, particularly owing to their relatively low values \citep{2022MNRAS.515..256P}. However, recent observational studies have inferred $\sim$\,$\rm{\mu G}$ magnetic field strengths in the CGM of a number of galaxies \citep{mao2017, prochaska2019, lan2020, mannings2022, osullivan2023, heesen2023}, which is of order of the field strengths found in the gaseous halos of galaxies at $z \sim 0$ in the IllustrisTNG simulations \citep{marinacci2018, nelson2018, ramesh2023a}. Such inferences are made possible through measurements of Faraday rotation measure (RM), the phenomenon by which the polarisation vector changes as light propagates through a region with non-zero magnetic field \citep[see e.g.][]{kim2016,rudnick2023,takahashi2023}.

Recently, \cite{heesen2023} used the LOFAR Two-metre Sky Survey \citep[LoTSS;][]{osullivan2023} to identify a trend of RM strength with galaxy azimuthal angle. Stacking a sample of $21$ unique galaxies ($29$ sightlines) at $z$\,$\sim$\,$0$, with impact parameters $\lesssim 100$~kpc, they find that RM values are $\sim$\,$2-3$ times higher close to the minor axis of galaxies in comparison to the major axis. Similar work was performed by {\color{blue}B\"ockmann et al. (2023, submitted)} who used RM values from the MIGHTEE-POL survey by MeerKAT in the XMM-LSS and COSMOS fields. However, they did not study the azimuthal dependence of the RM. Still, they find a central RM excess of $5.6\pm 2.3$\,$\rm rad\,m^{-2}$ with $2.5\sigma$ significance around star-forming galaxies with a median redshift of $z = 0.46$ for impact parameters below 130\,kpc. They find no evidence for a correlation between RM and redshift. They conclude that  mostly luminous, star-forming galaxies with impact parameters of $<$ 130 kpc contribute to the RM of the distant radio sources.

In this paper, we use the cosmological magnetohydrodynamical simulation TNG50 to explore the angular structure of magnetic fields in the CGM, in order to provide theoretical interpretation for these RM observations, while also making predictions for future observational surveys. The paper is organised as follows: in Section~\ref{methods}, we describe the TNG50 simulation and the methods we utilise throughout the paper. We present and discuss our results in Section~\ref{results}, and summarise our findings in Section~\ref{summary}.

\section{Methods}\label{methods}

\begin{figure*}
\centering 
\includegraphics[width=0.99\textwidth]{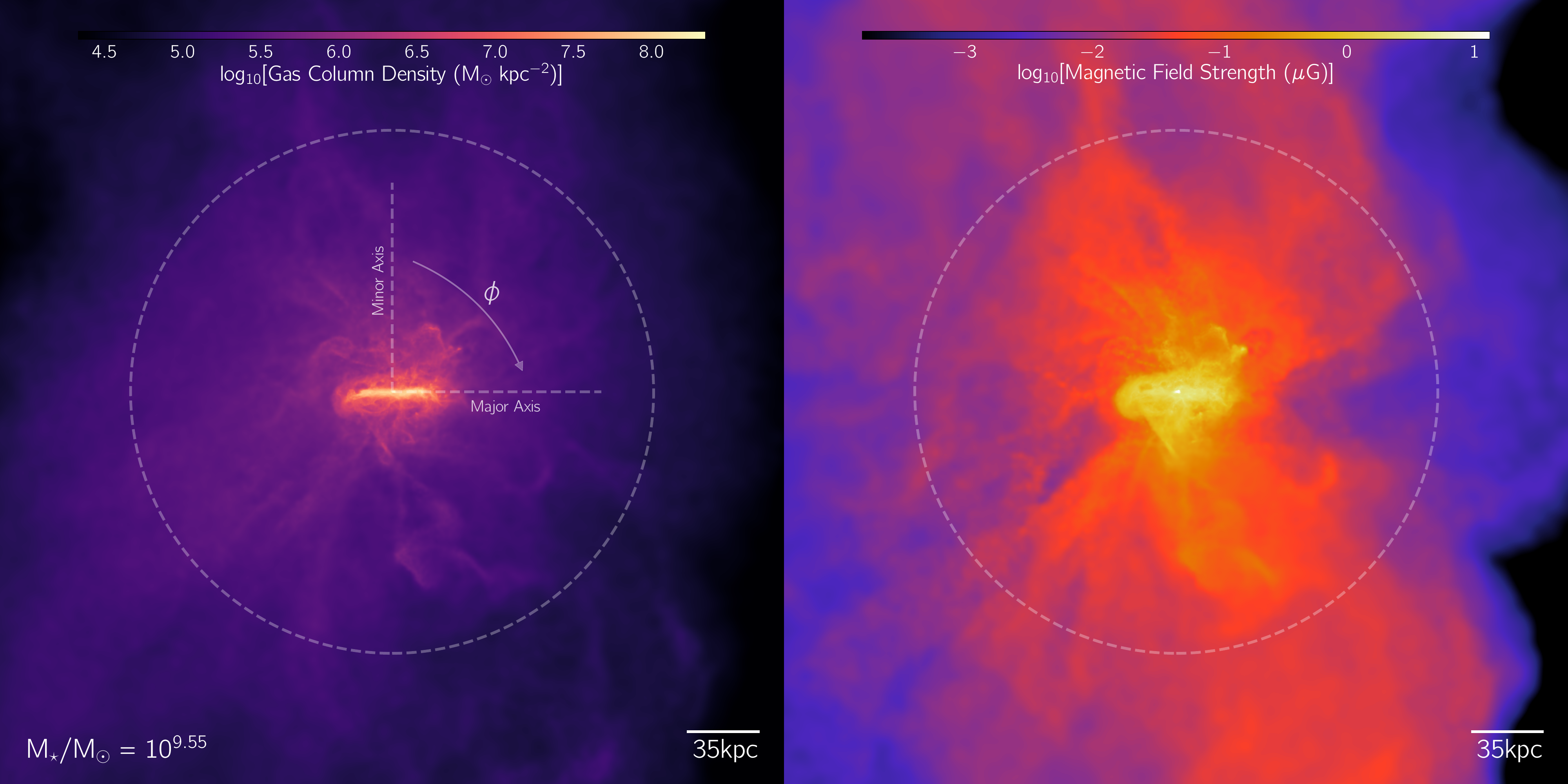}
\caption{Visualisation of a random TNG50 $\rm{M_\star} \sim 10^{9.55} \rm{M_\odot}$ galaxy at $z=0$ rotated edge-on. The left panel shows the gas column density along a projection that extends $\pm 1.5~R_{\rm 200c}$ in the line-of-sight direction, while the right panel shows a mass-weighted projection of magnetic field strength. In both panels, the dashed circles show the virial radius of the halo. At this mass range, except in the innermost regions of the halo, magnetic field strengths are preferentially stronger closer to the minor axis, i.e. at low values of $\phi$, defined as the angle with respect to the minor axis, as shown in the left panel.}
\label{fig:visFigure}
\end{figure*}

\subsection{The TNG50 simulation}\label{TNG}

For our analysis we use the TNG50-1 simulation \citep[hereafter, TNG50;][]{pillepich2019, nelson2019} of the IllustrisTNG project \citep{springel2018, naiman2018, pillepich2018b, marinacci2018, nelson2018}, a series of cosmological magnetohydrodynamical simulations of galaxy formation. This is the highest resolution TNG simulation, with an average baryonic mass resolution of $\sim 8 \times 10^4$ M$_\odot$ within a volume of $\sim (50 \rm{cMpc})^3$. It was run with the moving mesh code \textsc{arepo} \citep{springel2010}, and the fiducial `TNG model' for galaxy formation physics \citep{weinberger2017, pillepich2018a}: this includes key processes such as primordial and metal-line cooling, heating from a metagalactic background radiation field, self-shielding of dense gas, star formation, stellar evolution and enrichment, tracking of supernovae type Ia, II, and AGB stars, supermassive black hole formation, accretion, merging, magnetic fields, and other physics that is expected to play an important role in the growth and evolution of galaxies.

Importantly, the TNG model includes both stellar and supermassive blackhole (SMBH) feedback processes. Stellar feedback transfers energy to gas in the interstellar medium through a decoupled wind scheme \citep{springel2003} to produce mass-loaded galactic-scale outflows \citep{pillepich2018a}. On the other hand, SMBH feedback operates in one of two modes depending on the accretion state of the SMBH. At high accretion rates, i.e. when the ratio of Bondi to Eddingtion rates is large, thermal energy is continuously deposited into neighbouring gas cells. At low accretion rates, feedback energy is instead imparted in the form of time-discrete, high-velocity, randomly oriented kinetic kicks \citep{weinberger2017}. Low-mass SMBHs are preferentially in the thermal mode, while more massive SMBHs are typically in the kinetic mode, with the transition occurring roughly at $M_\star \sim 10^{10.5} \rm{M}_\odot$ \citep[corresponding to $M\rm{_{BH}} \sim 10^8 \rm{M_\odot}$;][]{nelson2018,weinberger2018}.

Importantly, for the structure and directionality of galactic-scale outflows, the TNG feedback processes always inject energy isotropically. For stellar winds and thermal input from SMBHs in the high-accretion state this is true by construction for each energy injection, while the kinetic energy inputs from SMBHs in the low-accretion state are randomly oriented, such that when time-averaged, energy injection in this mode is also isotropic. Nonetheless, for both low-mass and high-mass galaxies, outflows in TNG naturally collimate and propagate along preferred directions \citep{nelson2019}.

The TNG simulations adopt a cosmology consistent with the Planck 2015 analysis \citep{planck2016}, with: $\Omega_\Lambda = 0.6911$, $\Omega_{\rm m} = 0.3089$, $\Omega_{\rm b} = 0.0486$ and $h = 0.6774$.

\subsection{Magnetic Fields in TNG}

A significant difference between TNG and other large-volume cosmological simulations is the inclusion of (ideal) MHD \citep{pakmor2011, pakmor2014}. At the start of the simulation ($z=127$), a uniform primordial field of $10^{-14}$ comoving Gauss is seeded. This field then becomes amplified as a natural result of structure formation and feedback processes, including exponential early-time growth due to small-scale dynamos within halos, followed by a slower linear amplification phase due to differential rotation within disk galaxies \citep{pakmor2017, pakmor2020}.

In the TNG simulations, magnetic field strengths within galaxies reach $1-10 \,\mu\rm{G}$ levels by $z \sim 0$, consistent with observational inferences \citep{marinacci2018}. Magnetic fields have complex topology and connections with galaxy and feedback processes. For instance, at the Milky Way-mass scale, $|B|$ strengths are systematically higher for blue versus red galaxies \citep{nelson2018}. Magnetic fields permeate the circumgalactic medium of Milky Way-like galaxies \citep{ramesh2023a}. Amplified magnetic fields are also ejected to large scales by feedback, producing over-magnetized bubbles around massive halos \citep{aramburo2021}. Finally, the presence of magnetic fields alters galaxy properties, scaling relations, and global statistics, including galaxy sizes, SMBH and stellar masses, halo-scale gas fractions, the $z=0$ stellar mass function, and the cosmic star formation rate density \citep{pillepich2018a}.

Throughout this work, in order to derive magnetic field strength as a function of projected distance away from galaxies, we compute two-dimensional maps of $|B|$ using mass-weighted projections and the standard SPH kernel, with the smoothing length scaled in accordance with the gas cell radius. Unless otherwise stated, the projection depth (i.e. along the line of sight direction) scales with the impact factor: for an impact factor $b$, all gas within $\pm b$ along the perpendicular direction is included. 

\subsection{Synthetic Faraday Rotation Measures}\label{FRM}

We calculate synthetic Faraday rotation measure (RM) values by integrating sightlines through the TNG50 simulation volume. At $z=0$, we generate $10,000$ primary sightlines around a galaxy sample spanning a range of stellar masses (Section~\ref{sec:rm}), randomly located in angle and distance from the galaxy center out to twice the virial radius. All sightlines are oriented along the $\hat{z}$-axis of the simulation domain, and thus random with respect to the orientations of all galaxies. In addition, we use eight discrete snapshots at redshifts $z=\{0.0, 0.1, 0.2, 0.3, 0.4, 0.5, 0.7, 1.0\}$ to generate $N = 2000 \times 2000 = 4 \times 10^{6}$ random background sightlines and propagate each for a total distance equal to the simulation box length of 35 cMpc/$h$. Background sightlines can then be randomly stacked together with primary sightlines such that the total line-of-sight distance (i.e. projection depth) is equal to the light travel distance to a given redshift $z_{\rm{bk}}$. In all cases we calculate the line-of-sight integral

\begin{equation}
{\rm RM} = 0.812 \int n_{\rm e}\, B_{||}\, (1+z)^{-2} dl \;\;\;\rm{[rad\, m^{-2}]}
\label{eq:FRM}
\end{equation}

\noindent by ray tracing through the gas distribution. We do so directly on its native representation of an unstructured Voronoi tessellation, computing the cell-by-cell intersections with each ray and the path lengths $dl$ (in $\rm{pc}$) through each cell (\textcolor{blue}{Nelson in prep}). The values of the line-of-sight component of the magnetic field, $B_{||}$ (in $\mu\rm{G}$), as well as the electron number density, $n_{\rm{e}}$ (in $\rm{cm^{-3}}$), are constant within each cell, and taken directly from the simulation output. The only exception is for star-forming gas, which is subject to an additional pressure term from our effective two-phase interstellar medium model \citep{springel2003}. For these star-forming gas cells, we decompose the mass into its cold and hot components, and take $n_{\rm{e}}$ of the hot (ionized) phase only (following \textcolor{blue}{Nelson et al. 2023}). This approach is most consistent with the gas state in the TNG model and was also used in previous RM studies from simulations with similar ISM models \citep{pakmor2020,mannings2022}.

\begin{figure*}
\centering 
\includegraphics[width=0.9\textwidth]{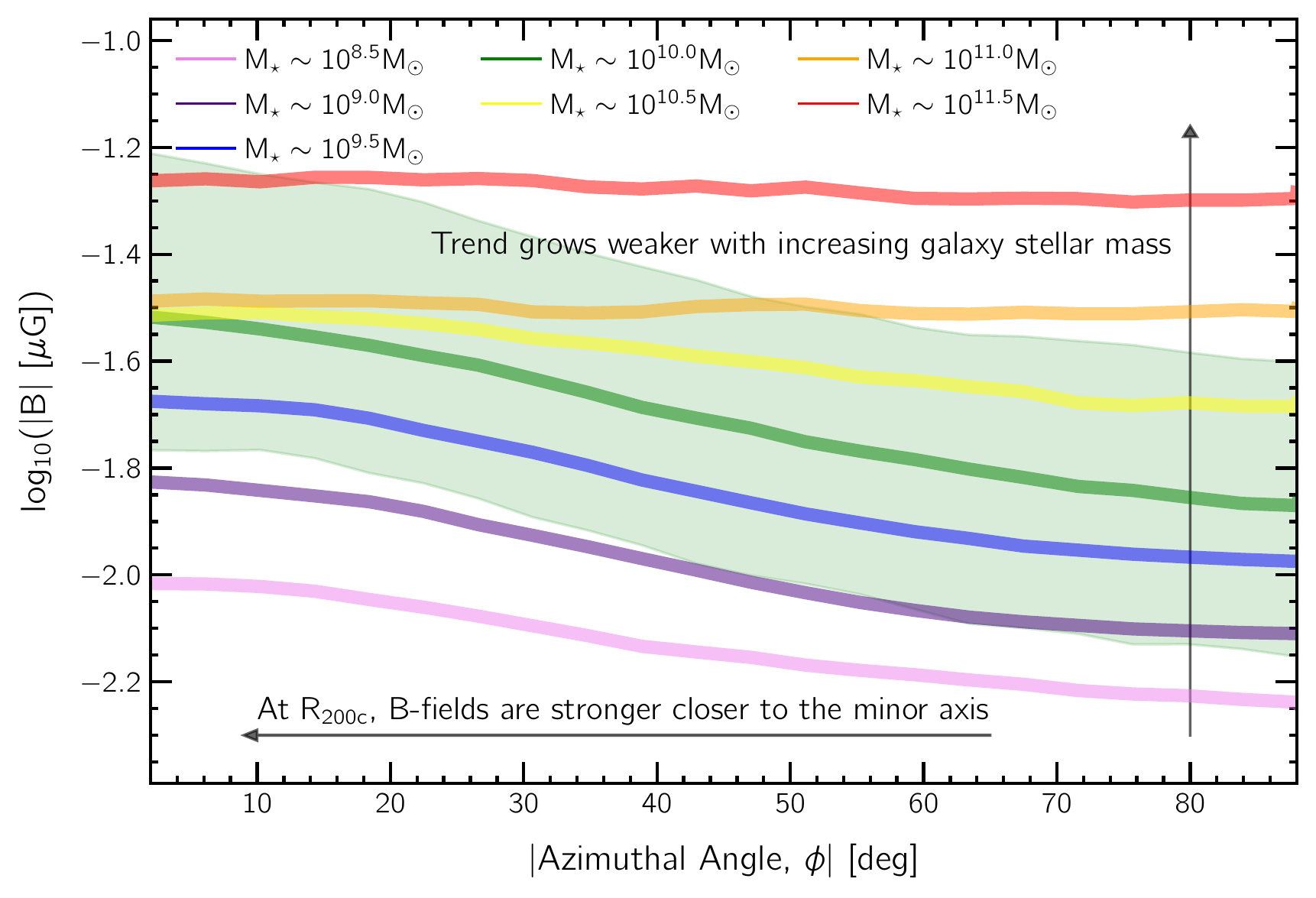}
\caption{Magnetic field strength as a function of azimuthal angle ($\phi$) at the virial radius, for galaxies in seven different mass bins from $10^{8.5} \lesssim M_\star / \rm{M}_\odot \lesssim 10^{11.5}$, from TNG50 at $z=0$. Individual curves show medians. The shaded region, corresponding to 16$^{\rm{th}}$-84$^{\rm{th}}$ percentile values, is shown only for one bin $M_\star \sim 10^{10}\,\rm{M}_\odot$ for visual clarity. The anisotropy is strongest for M$_\star$\,$\sim$\,$10^{10}\,\rm{M_\odot}$ galaxies, and grows weaker for more massive galaxies.}
\label{fig:b_vs_angle_massBins}
\end{figure*}

For the primary RM rays, we focus on galaxies that are oriented roughly edge-on in projection along the $\hat{z}$-axis of the simulation box, allowing us to define an azimuthal angle as in observations. We do so by restricting our analysis to galaxies with an ellipticity parameter $e \gtrsim 0.75$, where $e = 1 - b/a$ and $a~(b)$ is the semi-major (semi-minor) axis length when viewed along this projection, as derived by \textsc{statmorph} morphological measurements in SDSS r-band synthetic stellar light images \citep{rodriguezgomez2019}. 

\subsection{Galaxy Sample and Definitions}

In this work we only select central galaxies, i.e. those that lie at the potential minimum of their respective dark matter halos, as determined by the \textsc{Subfind} substructure identification algorithm \citep{springel2001}. Throughout this work, $\rm{M_\star}$ refers to the stellar mass within twice the stellar half mass radius.

When considering true magnetic field strengths (and not RM sightlines), in order to connect to galaxy orientation and derive a well-defined azimuthal angle, we rotate galaxies to be edge-on using a diagonalization of the moment of inertia tensor. 

\section{Results}\label{results}

In Figure~\ref{fig:visFigure}, we begin with a visualisation of a gaseous halo around a ($z=0$) TNG50 $\rm{M_\star} \sim 10^{9.55} \rm{M_\odot}$ galaxy. The image extends $\pm 1.5~R_{\rm 200c}$ from edge-to-edge and in the projection direction, and shows the central galaxy oriented edge-on. In both panels, the dashed circle shows the virial radius, $R_{\rm 200c}$, of the parent halo.

In the left panel, we show the gas column density. Values are largest along the disk major axis direction, and decrease with increasing distance from the disk plane. The panel also illustrates the definition we follow for the azimuthal angle ($\phi$), which we define with respect to the minor axis, i.e. the direction perpendicular to the disk when viewed in this edge-on projection. A value of $\phi=0^\circ$ thus corresponds to sightlines along the minor axis, while sightlines along the major axis are situated at $\phi=90^\circ$.

The panel on the right shows a mass-weighted projection of magnetic field strength for the same galaxy, in the same projection. Magnetic fields are strongest along the disk, owing to the relatively large densities \citep{dolag2005}. As a result, at small impact parameters ($\lesssim 0.25~R_{\rm 200c}$), sightlines closer to the major axis ($\phi \sim 90^\circ$) are more magnetized than those at lower azimuthal angles, where the densities are typically lower than that of the disk. However, at larger galactocentric distances, sightlines at lower azimuthal angles are more magnetized versus higher $\phi$ counterparts. Such a trend is also visible in gas density (left panel), although the level of anisotropy is weaker.

In Figure~\ref{fig:b_vs_angle_massBins}, we quantify this angular anisotropy in magnetic field strengths seen visually above. In seven different colors, we show the median magnetic field strength as a function of azimuthal angle for galaxies stacked in seven different mass bins (of width $\pm 0.2$~dex) spanning $10^{8.5} \lesssim M_\star / \rm{M}_\odot \lesssim 10^{11.5}$. The number of galaxies in each bin varies from a few ten to a few hundred, with the lower stellar mass bins comprising of more objects. In each case, we stack sightlines with impact parameters in the range $[0.95, 1.05]~R_{\rm 200c}$, i.e. around the virial radius of the halo. For low-mass galaxies ($M_\star \lesssim 10^{10.5} \rm{M}_\odot$), a clear angular signal is visible, wherein B-field strengths are higher by $0.25-0.35$~dex along the minor axis ($\phi=0$) with respect to the major axis ($\phi=90$). A strong stellar mass trend exists. Above $M_\star$\,$\sim$\,$10^{10.0}~\rm{M}_\odot$, the anisotropy weakens with increasing stellar mass, with the difference between the minor and major axes eventually decreasing to $\sim 0$ for the highest stellar mass bin.

In the shaded region, we show the 16$^{\rm{th}}$-84$^{\rm{th}}$ percentile values, albeit only for one bin ($M_\star$\,$\sim$\,$10^{10.0}~\rm{M}_\odot$) for visual clarity. The percentile regions are relatively broad, stretching $\sim \pm 0.25$~dex across the median at all azimuthal angles. As discussed in \cite{peroux2020}, this poses a challenge to observational studies: if only a handful of sightlines are available for galaxies of a given stellar mass bin, randomly distributed in azimuthal angle, it is possible that such a trend may be missed due to the large scatter.

\begin{figure}
\centering 
\includegraphics[width=0.45\textwidth]{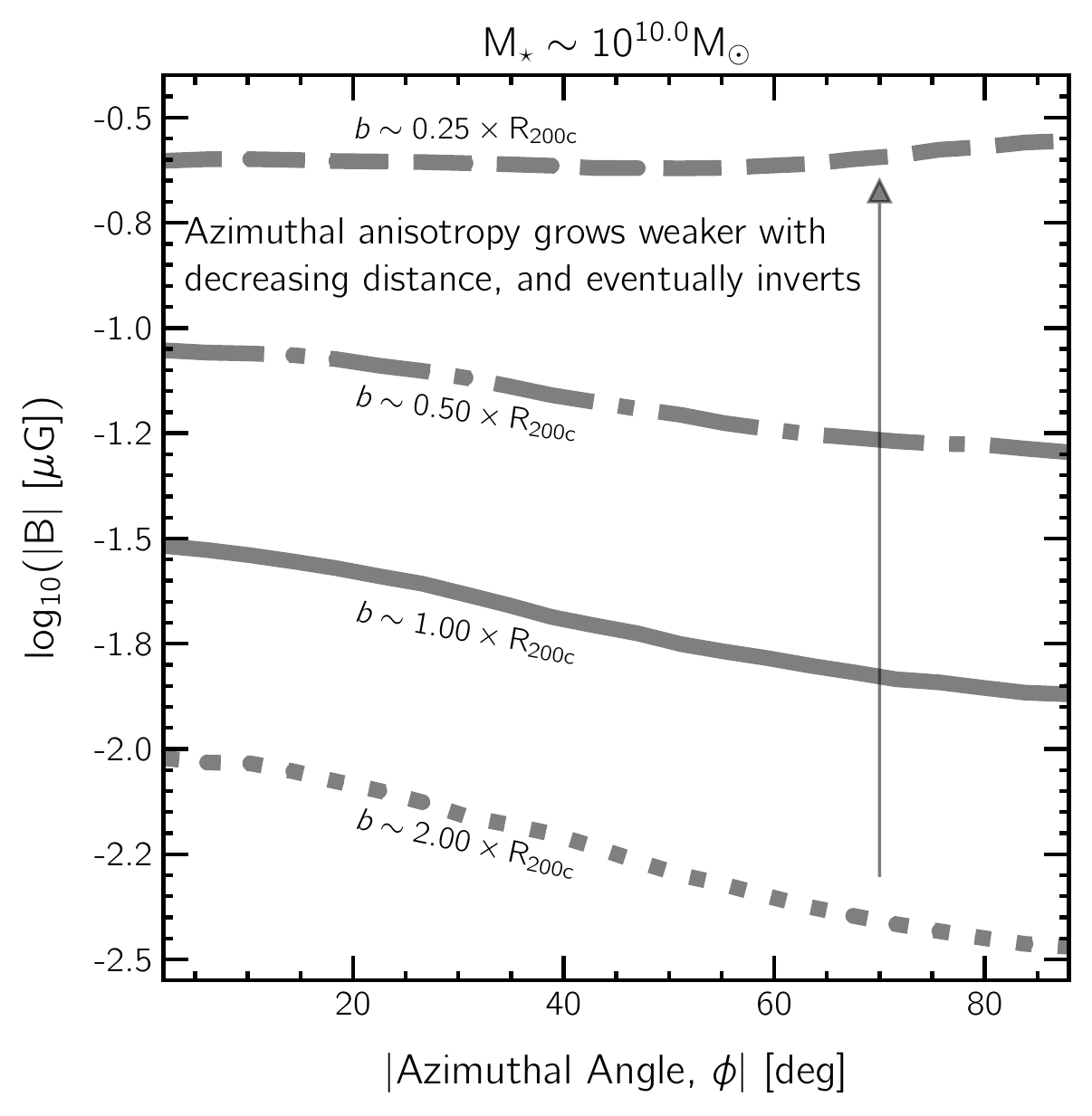}
\caption{Magnetic field strength as a function of azimuthal angle ($\phi$) at different impact parameters, for galaxies with stellar mass $\sim 10^{10} \rm{M_\odot}$. The angular anisotropy grows weaker with decreasing distance, and eventually inverts close to the disk ($\lesssim 0.25 \times \rm{R_{200c}}$).}
\label{fig:b_vs_angle_distanceTrends}
\end{figure}

In Figure~\ref{fig:b_vs_angle_distanceTrends}, we show the trend with galactocentric distance. We consider galaxies with $M_\star$\,$\sim$\,$10^{10.0}\,\rm{M}_\odot$ and stack sightlines at four different impact parameters: $0.25 R_{\rm 200c}$ (dashed curves), $0.50 R_{\rm 200c}$ (dash-dotted), $R_{\rm 200c}$ (solid; same as Figure~\ref{fig:b_vs_angle_massBins}), and $2 R_{\rm 200c}$ (dotted). 

At the smallest impact parameters, the median trend differs qualitatively from Figure~\ref{fig:b_vs_angle_massBins}. As discussed above, sightlines at such small impact parameters intersect the extended gaseous disk, where the density (and hence magnetic field strength) is higher at larger $\phi$, i.e. closer to the major axis. At larger impact parameters, the trend inverts, and the anisotropy is then stronger at larger galactocentric distances: a median difference between the minor and major axes of $\sim~0.2$~dex at $0.50 R_{\rm 200c}$ increases to $\sim~0.35$~dex at $R_{\rm 200c}$, and further to $\sim~0.45$~dex at $2 R_{\rm 200c}$. At this mass range, the anisotropy persists up to impact parameters as large as $\sim$\,$10 R_{\rm 200c}$, although the field strengths at such distances are much smaller ($\lesssim$\,$10^{-3.5}~\mu$G) in comparison to the halo. While not shown explicitly, we mention that all trends of Figure~\ref{fig:b_vs_angle_distanceTrends} are qualitatively similar for galaxies in other stellar mass bins.

We note that these trends of magnetic field strength versus azimuthal angle in the CGM are qualitatively similar to angular anisotropies of other gas properties predicted by recent cosmological hydrodynamical simulations including TNG \citep{peroux2020, truong2021, yang2023}. In particular, the direction of the effect is consistent with the angular modulation of CGM gas density \citep{truong2021}, as expected.

\begin{figure}
\centering 
\includegraphics[width=0.45\textwidth]{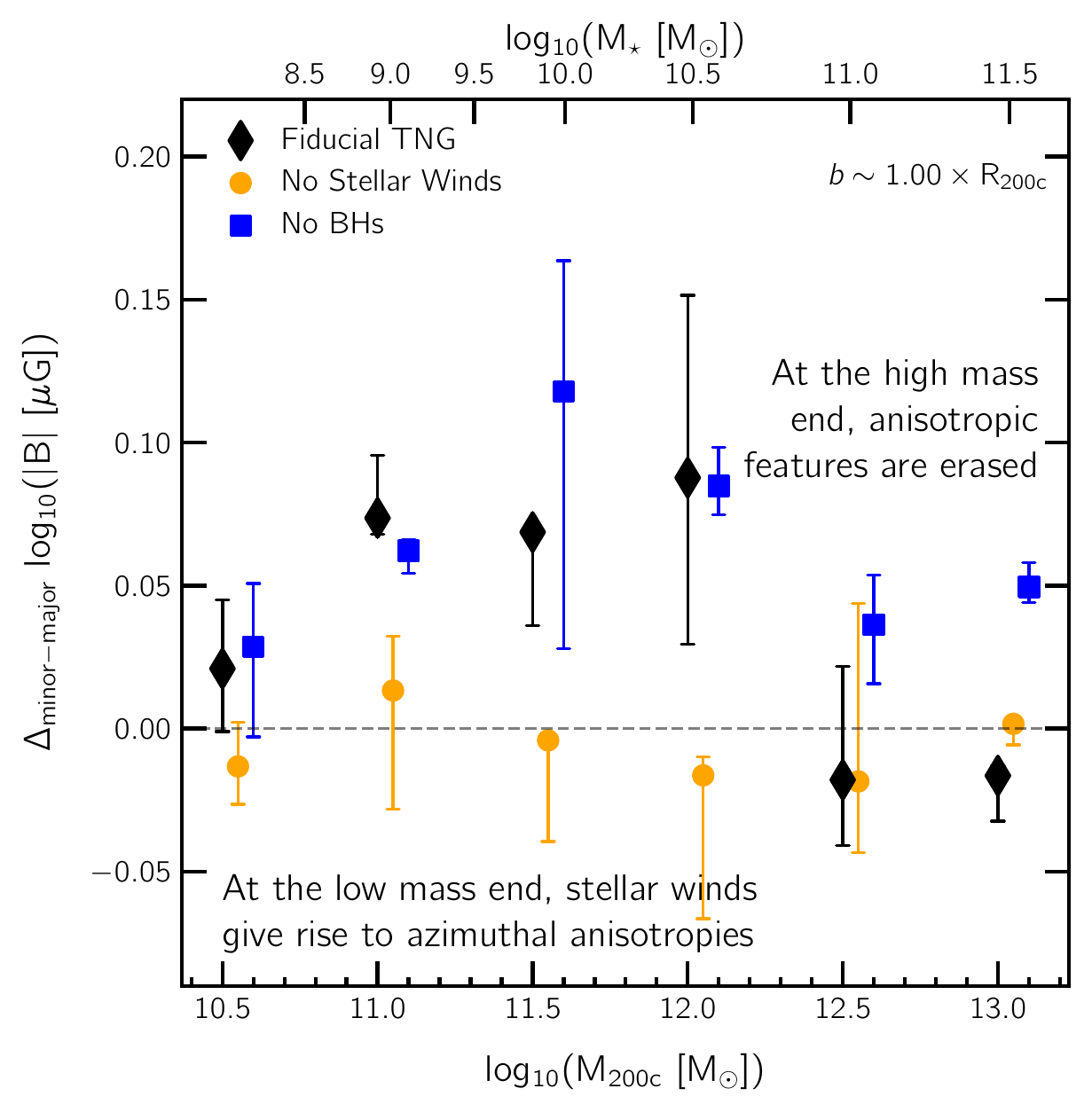}
\caption{Difference in B-field strengths between minor/major axes at R$_{\rm{200c}}$, as a function of halo mass, for different TNG model variation runs, with the fiducial model in black (see text). For reference we include the mean stellar mass as a function of halo mass on the top x-axis, from TNG50. Stellar feedback processes launch dense, magnetized gas perpendicular to the disk, giving rise to angular anisotropies at the low stellar mass end (M$_\star$\,$\lesssim$\,$10^{10.5} \rm{M_\odot}$). Without stellar feedback (orange), the anistropy is not present. For more massive galaxies, the kinetic mode of SMBH feedback dominates and results in galaxy quenching. The resulting outflows act to erase the anistropy of CGM magnetic fields at this mass scale, which is otherwise still present (blue).}
\label{fig:b_vs_angle_variationTest}
\end{figure}

In Figure~\ref{fig:b_vs_angle_variationTest}, we identify the physical processes responsible for producing the anisotropy in B-fields discussed so far. To do so, we use the TNG variations, a large set of simulations run with perturbations to the fiducial model \citep[as discussed and first presented in][]{pillepich2018a}. These use a smaller box of $L_{\rm{box}} \sim 37$~Mpc and have a resolution equal to that of TNG100-1 ($m_{\rm{b}} \sim 2 \times 10^{6}~\rm{M_\odot}$). Specifically, we consider three variants: (a) Fiducial TNG: the baseline model with all aspects unchanged, (b) No Stellar Winds: excluding resolve stellar feedback processes i.e. the stellar feedback-driven galactic winds, and (c) No BHs: the TNG model, but with no black holes seeded, and hence no black hole feedback.

To quantify the anisotropy between the minor and major axes, we stack all sightlines into two bins of azimuthal angle: $[0,45]^\circ$ and $[45,90]^\circ$. All sightlines in the former bin are associated with the minor axis, and the latter with the major axis. We focus on the difference between these two bins, labelled as $\Delta_{\rm{minor-major}}$ on the y-axis, as a function of halo virial mass $M_{\rm 200c}$. For reference, the median stellar mass from TNG50 is shown on the top x-axis. Symbols correspond to median differences, while error bars show the 16$^{\rm{th}}$-84$^{\rm{th}}$ percentile values. Results from the fiducial, no stellar winds, and no BHs variation boxes are shown in black, orange, and blue, respectively. In this plot, we only select sightlines with impact parameters close to the virial radius. Although not shown explicitly, the trends discussed here are qualitatively similar at other impact parameters as well. 

For lower halo masses $M_{\rm 200c}$\,$\lesssim$\,$10^{12}$\,$\rm{M_\odot}$, corresponding to $M_\star$\,$\lesssim$\,$10^{10.5}$\,$\rm{M_\odot}$ in TNG50, values of $\Delta_{\rm{minor-major}}$ are more or less consistent between the Fiducial and No BHs runs, and are positive, i.e. in this mass regime, angular anisotropies are present, and are largely unaffected by the activity of the central SMBH. However, when stellar winds are turned off (orange points), the anisotropy disappears, with $\Delta_{\rm{minor-major}}$ oscillating randomly around a value of zero. This suggests that outflows driven by stellar winds launch over-dense, magnetised gas preferentially perpendicular to the disk, and are largely responsible for the angular structure of magnetic fields. As noted in earlier studies \citep[e.g.][]{nelson2019}, outflows prefer the direction perpendicular to the disk since densities along these paths are lower in comparison to that of the disk, and a path of least resistance is established along the pressure gradient.

For more massive halos, i.e. $M_{\rm 200c}$\,$\gtrsim$\,$10^{12.5}~\rm{M_\odot}$, corresponding to $M_\star$\,$\gtrsim$\,$10^{11.0}$\,$\rm{M_\odot}$ in TNG50, the picture begins to change: at this mass scale, differences arise between the Fiducial and No BHs cases (black vs blue points), while the Fiducial and No Stellar winds cases (black vs orange points) are consistent with each other. Specifically, strong anisotropies between the minor and major axes cease to exist ($\Delta_{\rm{minor-major}}$\,$\sim$\,$0$) when black holes are present. As discussed in \cite{truong2021}, this is likely due to the quenching of central galaxies by the kinetic mode of SMBH feedback: as galaxies become quenched, a well defined disk-like structure ceases to exist, due to which outflows no longer have a preferred direction. As a result, the kinetic kicks distribute energy uniformly in all directions, suppressing any angular structure previously present. 

While no other large volume cosmological simulation beyond TNG currently include the effects of magnetic fields, a few zoom-in projects do. This enables the study of the impact of MHD on the evolution of cosmic gas. For instance, the Auriga simulations produce magnetic field strengths and radial profiles which are overall similar to those in TNG, albeit only for Milky Way-like galaxies \citep{pakmor2020}. The FIRE simulations include variations with MHD \citep[e.g.][]{ponnada2022}, and previous studies have shown that field strengths at $z=0$ are generally weaker than in TNG \citep[see][and discussion therein]{ramesh2023a}. In both cases, although the angular anisotropy of magnetic fields has not yet been studied, given their origin in the collimation of galactic-scale outflows, we speculate that such signals may be present as well.\footnote{Note that the majority of the FIRE simulations run so far do not include black holes, and hence only the impact of stellar feedback would be visible.}

\subsection{The Observable: Rotation Measure}\label{sec:rm}

\begin{figure}
\centering 
\includegraphics[width=0.46\textwidth]{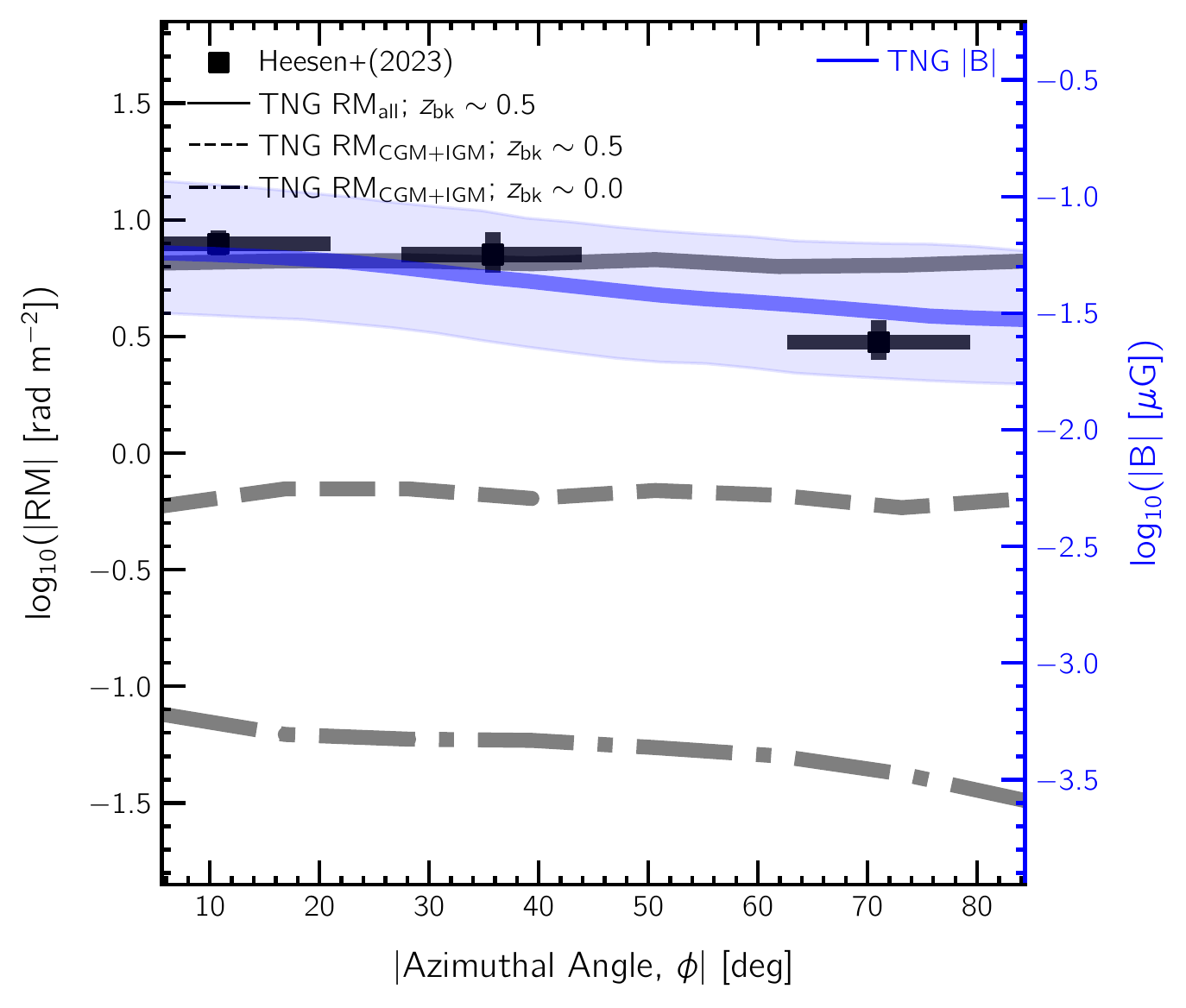}
\caption{Rotation measure as a function of azimuthal angle ($\phi$) at $\sim100$~kpc, for galaxies with stellar mass $\sim 10^{10}\, \rm{M_\odot}$. Black points show results from \protect\cite{heesen2023}, while the three gray lines correspond to three measurements from TNG, as elaborated in the main text. The blue curve shows the true median three dimensional magnetic field strength, for comparison.}
\label{fig:rm_v_angle}
\end{figure}

In Figure~\ref{fig:rm_v_angle}, we compare the results of the TNG50 simulation to observations, where we show the rotation measure (RM) as a function of azimuthal angle (left y-axis and gray lines). The black points correspond to those from \cite{heesen2023}, for impact parameters $b<100$~kpc. The three gray curves correspond to three different measurements using TNG for a selection of 70 galaxies with stellar masses M$_\star$\,$\sim$\,$10^{10}\,\rm{M_\odot}$, the typical stellar mass of objects in the sample of \cite{heesen2023}. The dash-dotted line only includes the $z=0~(\rm{d_{LoS}}$\,$\sim$\,$50$~Mpc) primary sightlines for each galaxy, i.e. the effect of the CGM and the local IGM (RM$_{\rm{CGM}}$ + RM$_{\rm{IGM}}$; $z_{\rm{bk}} \sim 0.0$). The dashed line includes additional background sightlines to yield a total line-of-sight distance out to $z\sim0.5$, i.e. the effect of the cosmological IGM is also included (RM$_{\rm{CGM}}$ + RM$_{\rm{IGM}}$; $z_{\rm{bk}} \sim 0.5$) (see Section~\ref{FRM} for details). This is the typical background polarized source redshift of the \cite{heesen2023} sample \citep[see also][]{hackstein2020}. In the TNG simulations, these contributions from the cosmological IGM can be non-negligible, especially at $z \lesssim 2$, as a result of star formation and SMBH driven outflows expelling magnetized gas into their local environments \citep{aramburo2023}. In addition, other intervening galaxies may impact the net RM signal, although this is expected to be rare in the TNG simulations: \cite{aramburo2023} find that a fraction $\sim 10^{-4}$ of sightlines are impacted by such contributions. However, some other studies suggest that contributions due to intervening galaxies may be more important, possibly yielding a non-Gaussian distribution of background RMs \citep[e.g.][]{shah2021}.

In addition to the intervening CGM and IGM, other contributions to the observed RM signals are present. In particular, from the Milky Way galaxy and halo (foreground), and from the host environment/galaxy of the polarized source itself (background). The former contribution is removed by subtracting a model for the Galactic rotation measure, i.e. the effect of gas in the Milky Way. However, this subtraction unavoidably introduces uncertainty. In addition, instrumental noise and other contaminating contributions to the RM may also be present \citep[see e.g.][]{hackstein2019}.

We collectively account for all these external contributions, which we label RM$_{\rm{ext}}$, by adding random Gaussian noise, with the mean ($\sim$\,5.68 rad m$^{-2}$) and standard deviation ($\sim$\,5.50 rad m$^{-2}$) computed from the \cite{heesen2023} RMs at impact parameters $b>500$~kpc \citep[see also][]{basu2018}. The solid black curve shows the resulting RM values, i.e. RM$_{\rm{all}}$ = RM$_{\rm{ext}}$ + RM$_{\rm{CGM}}$ + RM$_{\rm{IGM}}$; $z_{\rm{bk}} \sim 0.5$. 

The dash-dotted curve (the intervening CGM and local IGM only) shows a drop of $\sim$\,$0.4$~dex from the minor to major axes, consistent with the $\sim$\,$0.4$~dex drop seen in the the black points. Note, however, that the RM excess, i.e. the difference between the values at the minor and major axes in linear scale, predicted by TNG ($\sim$\,$0.05$ rad m$^{-2}$) is about two orders of magnitude smaller than that of \cite{heesen2023} ($\sim$\,$4$ rad m$^{-2}$). While the relative anisotropy predicted by TNG is consistent with that of \cite{heesen2023}, the absolute excess is thus smaller (see discussion below).

However, when the effect of the cosmological IGM is included (dashed line), or once external effects are accounted for (solid line), the trend is buried in these non-local signals. This is because the true signal (RM$_{\rm{CGM}}$) predicted by TNG is almost always weaker than the other contributions (RM$_{\rm{ext}}$ and RM$_{\rm{IGM}}$; $z_{\rm{bk}} \sim 0.5$) along the line of sight \citep[see also][]{basu2018}. 

\begin{figure*}
\centering 
\includegraphics[width=0.95\textwidth]{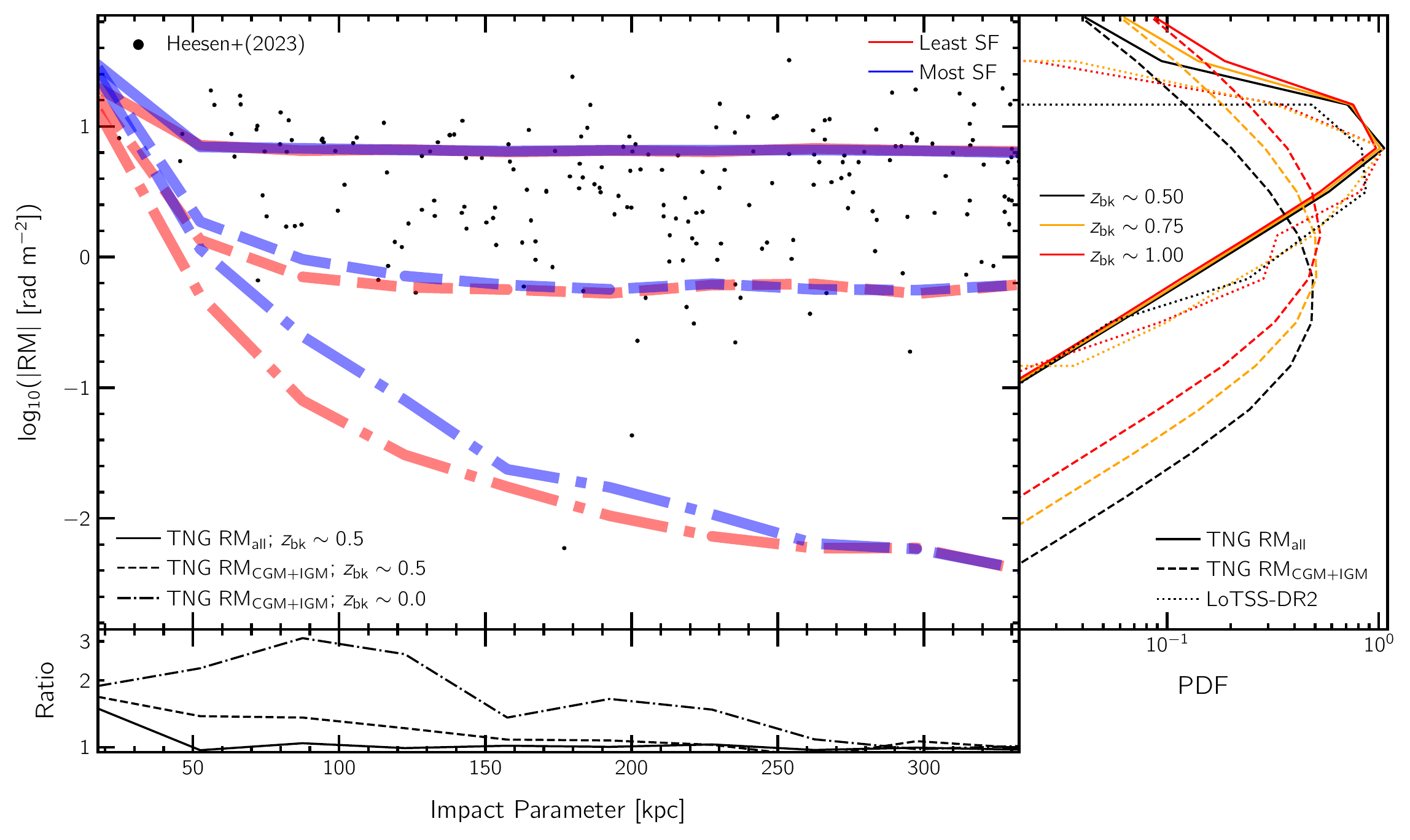}
\caption{In the main panel, we show the trend of rotation measure (RM) as a function of impact parameter for galaxies with M$_\star$\,$\sim$\,$10^{10}$\,$\rm{M_\odot}$. The sample is split into two bins based on sSFR values: $<10^{\rm{th}}$ percentile (least SF; red) and $>90^{\rm{th}}$ percentile (most SF; blue). Solid, dashed and dot-dashed curves correspond to three different synthetic TNG RM measurements, as elaborated in the main text. The ratio of the blue-to-red curves is shown in the bottom panel. Black dots show data points from \protect\cite{heesen2023}. In the right panel, we show PDFs of RM for random sightlines from TNG (solid and dashed curves), and those from the LoTSS-DR2 catalog \protect\citep[dotted curves;][]{osullivan2023}. Different colors correspond to different projection depths (refer to main text for details).}
\label{fig:rm_v_b}
\end{figure*}

In blue, for the purpose of comparison, we overlay the true (three dimensional) magnetic field strength-azimuthal angle median relation of these galaxies, i.e. the two y-axes are not related and have an arbitrary relationship for visualization only. The true magnetic field trend shows a clear decrease of $\sim$\,$0.4$~dex between the minor versus major axes directions. At the same time, the slope at large azimuthal angles $\phi \gtrsim 45^\circ$ is less steep (in $\log(B)$) than the observations (in $\log(\rm{RM})$), although the 16$^{\rm{th}}$-84$^{\rm{th}}$ percentile (shaded) region is broad. TNG is therefore in qualitative agreement with the results of \cite{heesen2023}, while also predicting that such trends are challenging to observe (see below). That is, the azimuthal angle dependence of RM is largely overtaken by external contributions of gas along the line of sight: either gas surrounding the polarised source, or IGM gas, in addition to any other sources of observational noise.

This is similar to challenges faced in the interpretation of dispersion measure (DM) values from observations of fast radio burst (FRBs). Like RMs, the observed DM values result from the combined contribution of the Milky Way, the IGM, the host, and the source. It is not straightforward to disentangle these components and retrieve the contribution of the source alone, owing to its weak amplitude in comparison to other components \citep[e.g.][]{zhang2021}.

In Figure~\ref{fig:rm_v_b}, we move away from angular structure, and focus on the broad trend of RM as a function of impact parameter. In the main panel we show three different synthetic values of RM for the same selection of 70 galaxies with stellar masses M$_\star$\,$\sim$\,$10^{10}$\,$\rm{M_\odot}$ as above. These are: RM$_{\rm{CGM}}$ + RM$_{\rm{IGM}}$; $z_{\rm{bk}} \sim 0.0$ (dot dashed), RM$_{\rm{CGM}}$ + RM$_{\rm{IGM}}$; $z_{\rm{bk}} \sim 0.5$  (dashed) and RM$_{\rm{all}}$ (solid). Further, we split the galaxies into two sub-samples based on their specific star formation rate (sSFR)\footnote{We define sSFR as the ratio of star formation rate (within twice the stellar half mass radius) and M$\star$.}. Galaxies with values less than the $<10^{\rm{th}}$ percentile (i.e. the seven least star forming in this sample; median sSFR $\sim$\,$10^{-10.7}$~yr$^{-1}$; red) are contrasted against the $>90^{\rm{th}}$ percentile (i.e. the seven most star forming; median sSFR $\sim$\,$10^{-9.8}$~yr$^{-1}$; blue). We therefore study whether star-forming versus quiescent galaxies have different RM amplitudes in their CGM, motivated by similar dichotomies present in the TNG simulations e.g. in halo OVI abundance \citep{nelson2018b} and in X-ray luminosity \citep{truong2020}, the latter potentially detected in eROSITA stacking \citep{comparat2022,chadayammuri2022}.

In the CGM plus local IGM only RMs (dot dashed curve), a difference between the blue versus red curves is apparent out to impact parameters of $\sim$\,$250$\,kpc, wherein galaxies with higher star formation rates have larger values of RM. We interpret this as higher SFR galaxies launching stronger galactic-scale outflows, driving more magnetized gas into the CGM. When the cosmological IGM is included, i.e. out to $z_{\rm{bk}} \sim 0.5$ (dashed curves), the differences between the two curves disappear beyond $\sim$\,$125$\,kpc. With the addition of RM$_{\rm{ext}}$ component, differences are only clear close to the galaxy ($\lesssim$\,$50$\,kpc). To quantify these differences between the two sub-samples, the bottom panel shows the ratio of the blue to red curves. The peak amplitude of this effect is a factor of $\sim 3$ larger RMs in high versus low SFR galaxies, which occurs at impact parameters of $\sim 100$\,kpc. The quantitative strength of this dichotomy undoubtedly varies with galaxy mass as well as redshift, but its qualitative existence shows that observables of the magnetic fields in the CGM encode information on galactic feedback processes.

In black dots, we show the full sample from \cite{heesen2023}. These observed galaxies have not been split by star formation rate. All but a few measurements are for impact parameters $\gtrsim$\,$50$\,kpc, and the vast majority are at much larger distances, as statistically expected for random background-foreground associations. Observational detection of the differential RM signal between the blue and red sub-samples would clearly be a challenge given the level of scatter, once again primarily as a result of external contributions by gas along the line of sight. Only at small impact parameters does the local CGM signal dominate, making small separation pairs extremely valuable. Future large surveys such as POSSUM on ASKAP \citep{gaensler2010}, with LOFAR \citep{osullivan2020}, and SKA-era instruments \citep{heald2020}, will enable this science. Note that the linearity of the solid curves at small impact parameters is a result of using large bins.

In the right panel, we show PDFs of RM values for several cases of interest which are \textit{not} related to intervening galaxies. For a selection of $10^6$ \textit{random} TNG sightlines, we show RM$_{\rm{all}}$ in solid curves, and RM$_{\rm{CGM}}$ + RM$_{\rm{IGM}}$ in dashed lines. The dotted lines are observational data from the LoTSS-DR2 catalog \citep{osullivan2023}. In each of these cases, we split the samples (i.e. select) three different values of $z_{\rm{bk}}$: 0.5 (black), 0.75 (orange) and 1.0 (red). Each has different path-lengths of the cosmological IGM contribution included. For the LoTSS catalog, we bin the spectroscopic redshifts of polarised sources around the three different redshift values.

In the RM$_{\rm{CGM}}$ + RM$_{\rm{IGM}}$ case (dashed), increasing the traversal depth through the IGM leads to a systematic increase in RM values. The black curve, corresponding to $z_{\rm{bk}} \sim 0.5$, peaks at a value of $\sim$\,$0.4$ rad m$^{-2}$. Increasing the projection depth to $z_{\rm{bk}} \sim 0.75$ shifts the PDF by $\sim$\,$0.4$ dex towards higher RM values, and further by $\sim$\,$0.4$ dex for $z_{\rm{bk}} \sim 1.0$. Integrating to greater distances thus yields a larger value of RM, indicating that large-scale contributions as sightlines traverse the magnetized cosmic web continue to accrue. This suggests that, even though the integral in Equation~\ref{eq:FRM} is sensitive to the sign of $B_{||}$, positive and negative values i.e. field reversals do not average out \citep[see also][]{vacca2016}.

The solid curves show the PDFs with the RM$_{\rm{ext}}$ included. In this case, the three PDFs peak at roughly the same value, with differences between them apparent only in the high-RM tails ($\gtrsim$\,$15$ rad m$^{-2}$). That is, in the presence of the RM$_{\rm{ext}}$ component, the effect of the cosmological IGM is sub-dominant and largely disappears, except at values where the IGM component is comparable to RM$_{\rm{ext}}$. The three LoTSS (dotted) curves are in broad agreement with TNG, i.e. independent of $z_{\rm{bk}}$ except at the high-RM end. In the presence of a RM$_{\rm{ext}}$ component obtained from \cite{heesen2023}, TNG thus predicts that extracting RM$_{\rm{IGM}}$ is a challenge, as for RM$_{\rm{CGM}}$.

While recent studies using Auriga \citep{pakmor2020} and FIRE \citep{ponnada2022} have quantified the radial profiles of RM using RM mocks \citep[see also][]{liu2022}, only the RM$_{\rm{CGM}}$ component is included. Direct comparison with these studies is hence not possible, but we mention here that the RM values of \cite{pakmor2020} are comparable to our RM$_{\rm{IGM}}$; $z_{\rm{bk}} \sim 0.0$ case, i.e. when only the local IGM is included, while FIRE predicts lower values of RM, much like the weaker magnetic field strength values.

\section{Discussion}\label{discussion}

Overall, we confirm that RM$_{\rm{CGM}}$ measurements contain rich information on halo magnetic fields. TNG50 predicts that the angular anisotropy signal of magnetic fields observed by \cite{heesen2023} is present over a large range of halo and stellar masses, and not just for the M$_\star$\,$\sim$\,$10^{10}$\,M$_\odot$ bin that \cite{heesen2023} mainly probed. Moreover, TNG50 also predicts that the observed RM values are sensitive to feedback processes within the galaxy. Observing such trends would require denser sampling and better sensitivity, likely to be made possible by future surveys including POSSUM on ASKAP, as well as the Square Kilometer Array (SKA). Such measurements will be instrumental in improving the feedback physics implemented in future cosmological hydrodynamical simulations. However, if the TNG50 simulation is realistic, then this suggests that CGM RMs are challenging to measure observationally, especially at large impact parameters, due to their low values.

While the RM predicted by TNG is about two orders of magnitude smaller than the LoTSS catalog \citep{osullivan2023}, the discrepancy is even larger in comparison to some other results \citep[e.g.][]{sobey2019,prochaska2019,lan2020,seta2021}. One possibility is that the RM values predicted by TNG50 are indeed too low, in which case the most likely culprit is insufficiently large late time magnetic field strengths. Although under-predicted RM values can be a result of simulated $n_{\rm e}$ values being too low, this is unlikely to dominate our results (Figure~\ref{fig:neHist}). Further, while magnetic fields could be missing some degree of large-scaling ordering and hence have smaller coherence lengths than reality, it is unlikely that this can improve the RM signal by over an order of magnitude (Figure~\ref{fig:rmTest}). 

If simulated field strengths are too low, this could point to missing small-scale physical processes which boost amplification, or simply insufficient numerical resolution to capture small-scale dynamos on sufficiently short timescales \citep{pakmor2017}. While distributions of field strengths are believed to be converged at TNG50-1 resolution \citep[e.g.][]{marinacci2015,ramesh2023d}, some studies suggest that the \cite{powell1999} divergence-cleaning scheme utilised by \textsc{arepo} may not be sufficiently accurate \citep[e.g.][]{hopkins2016}, although \cite{pakmor2013} show that the results produced by \textsc{arepo} do not differ significantly in comparison to constrained transport schemes. \cite{carretti2023} recently showed that a uniform, homogeneous primordial magnetic field, like the one assumed in TNG, may not be an accurate representation of the universe, although various studies suggest that the exact nature of the seeding does not have a significant impact \citep[e.g.][]{pakmor2013,marinacci2015,vazza2017,martinalvarez2021}. Nevertheless, the TNG model can certainly be improved by incorporating more sophisticated seed fielding processes \citep[e.g.][]{garaldi2021}.

Finally, the apparent tension of RM could be due to complexities of the data analysis and comparison itself. Most clearly, observed values include the combined line-of-sight contributions from many sources other than the extragalactic CGM of the target galaxy. These encompass the local hot interstellar medium, the Milky Way halo itself, the cosmological IGM, and the intrinsic source. Many of these terms are highly uncertain. For example, the exceptionally high time variability of some repeating FRBs suggests that the source contribution, its subtraction to obtain RM$_{\rm{CGM}}$ for a given line-of-sight, and thus any comparison with cosmological simulations inherits this non-negligible uncertainty \citep{hilmarsson2021}.

The simulation outcomes will also change with the inclusion of more complex physics such as non-ideal MHD, or with the inclusion of cosmic rays which is believed to alter the flow structure of CGM gas \citep{butsky2022}. This could potentially yield larger RM and magnetic field strengths in the CGM, enhancing angular anisotropies in extragalactic halos to a level similar to that observed by \cite{heesen2023}. Moreover, differences in magnetic field properties between cosmological simulations run with different codes suggest that the numerical techniques employed to solve the equations of magnetohydrodynamics may still play a significant role \citep{ramesh2023a}. Constraints from observational data will help constrain magnetic fields in future cosmological magnetohydrodynamical simulations.

\section{Summary and Conclusions}\label{summary}

In this paper, we have shown that magnetic fields in the circumgalactic medium (CGM) of TNG50 galaxies have significant angular structure, such that gas along the minor axes of galaxies is more strongly magnetized with respect to the major axes. Our main findings can be summarised as follows:

\begin{enumerate}

\item At the virial radius, these angular anisotropies are strongest for galaxies with stellar masses M$_\star$\,$\sim$\,$10^{10}\,\rm{M_\odot}$, decreasing in strength by $\sim 0.35$~dex between the two galactic axes. The anisotropy grows weaker for both lower and higher mass galaxies (Figure~\ref{fig:b_vs_angle_massBins}).

\item A trend with distance is also present: the strength of the anisotropy weakens with decreasing distance, and for very small impact parameters ($\lesssim 0.25 \times \rm{R_{200c}}$), the direction of anisotropy eventually inverts, i.e gas along the major axis is more strongly magnetized, as a result of higher densities along the disk (Figure~\ref{fig:b_vs_angle_distanceTrends}).

\item This anisotropy arises as a result of two different feedback processes. At lower galaxy masses, stellar feedback launches dense, magnetized gas perpendicular to the disk plane of galaxies, yielding higher B-field strengths along the direction of the minor axis. This mode of feedback dominates for galaxies with M$_\star$\,$\lesssim$\,$10^{10.5} \rm{M_\odot}$. For more massive galaxies, SMBH (AGN) feedback quenches galaxies, washing out this angular structure as strong outflows produce underdense bubbles, while outflows also lose their directional collimation in high-mass galaxies lacking gaseous disks (Figure~\ref{fig:b_vs_angle_variationTest}).

\item We make predictions for Faraday Rotation Measure (RM) studies. TNG shows a qualitatively similar dependence of RM on galactic azimuthal angle as seen observationally \citep{heesen2023}. However, TNG also predicts small RM amplitudes that are difficult to observe above external contributions from gas present along the line of sight between the observer and the polarised source (Figure~\ref{fig:rm_v_angle}). In order to reproduce observed RM values, the magnetic field would have to be stronger. 

\item The profile of RM as a function of distance depends on galaxy star forming rate (SFR), with greater RM values around galaxies with higher SFR. This dichotomy reflects overall stronger magnetic fields in the circumgalactic medium of high SFR galaxies in comparison to their lower SFR and quenched counterparts. However, TNG50 predicts that these differences will be challenging to observe owing to their relatively low magnitudes, except at small impact parameters (Figure~\ref{fig:rm_v_b}).

\end{enumerate}

With the inclusion of magnetic fields, the TNG simulations provide a unique starting point to study the amplification, late-time structure, and impact of magnetic fields in/on galaxies and their gaseous halos. In the future, hydrodynamical simulations with substantially improved resolution in the circumgalactic medium, together with more sophisticated models for stellar and SMBH feedback, as well as currently absent physics including cosmic rays, are needed to fully capture the role of magnetic fields in galaxies and their halos.

\section*{Data Availability}

The IllustrisTNG simulations, including TNG50, are publicly available at \url{www.tng-project.org/data} \citep{nelson2019b}. Other data related to this publication is available upon request. 

\section*{Acknowledgements}

RR and DN acknowledge funding from the Deutsche Forschungsgemeinschaft (DFG) through an Emmy Noether Research Group (grant number NE 2441/1-1). RR is a Fellow of the International Max Planck Research School for Astronomy and Cosmic Physics at the University of Heidelberg (IMPRS-HD). MB acknowledges support from the Deutsche Forschungsgemeinschaft under Germany's Excellence Strategy - EXC 2121 "Quantum Universe" - 390833306. The authors thank the anonymous referee for constructive feedback that has greatly helped improve the quality of this work. The TNG50 simulation was run with compute time granted by the Gauss Centre for Supercomputing (GCS) under Large-Scale Projects GCS-DWAR on the GCS share of the supercomputer Hazel Hen at the High Performance Computing Center Stuttgart (HLRS). This analysis has been carried out on the VERA supercomputer of the Max Planck Institute for Astronomy (MPIA), operated by the Max Planck Computational Data Facility (MPCDF).

\bibliographystyle{mnras}
\bibliography{references}

\begin{thebibliography}{}
\makeatletter
\relax
\def\mn@urlcharsother{\let\do\@makeother \do\$\do\&\do\#\do\^\do\_\do\%\do\~}
\def\mn@doi{\begingroup\mn@urlcharsother \@ifnextchar [ {\mn@doi@}
  {\mn@doi@[]}}
\def\mn@doi@[#1]#2{\def\@tempa{#1}\ifx\@tempa\@empty \href
  {http://dx.doi.org/#2} {doi:#2}\else \href {http://dx.doi.org/#2} {#1}\fi
  \endgroup}
\def\mn@eprint#1#2{\mn@eprint@#1:#2::\@nil}
\def\mn@eprint@arXiv#1{\href {http://arxiv.org/abs/#1} {{\tt arXiv:#1}}}
\def\mn@eprint@dblp#1{\href {http://dblp.uni-trier.de/rec/bibtex/#1.xml}
  {dblp:#1}}
\def\mn@eprint@#1:#2:#3:#4\@nil{\def\@tempa {#1}\def\@tempb {#2}\def\@tempc
  {#3}\ifx \@tempc \@empty \let \@tempc \@tempb \let \@tempb \@tempa \fi \ifx
  \@tempb \@empty \def\@tempb {arXiv}\fi \@ifundefined
  {mn@eprint@\@tempb}{\@tempb:\@tempc}{\expandafter \expandafter \csname
  mn@eprint@\@tempb\endcsname \expandafter{\@tempc}}}

\bibitem[\protect\citeauthoryear{{Ar{\'a}mburo-Garc{\'\i}a}, {Bondarenko},
  {Boyarsky}, {Nelson}, {Pillepich}  \& {Sokolenko}}{{Ar{\'a}mburo-Garc{\'\i}a}
  et~al.}{2021}]{aramburo2021}
{Ar{\'a}mburo-Garc{\'\i}a} A.,  {Bondarenko} K.,  {Boyarsky} A.,  {Nelson} D.,
  {Pillepich} A.,   {Sokolenko} A.,  2021, \mn@doi [\mnras]
  {10.1093/mnras/stab1632}, \href
  {https://ui.adsabs.harvard.edu/abs/2021MNRAS.505.5038A} {505, 5038}

\bibitem[\protect\citeauthoryear{{Ar{\'a}mburo-Garc{\'\i}a}, {Bondarenko},
  {Boyarsky}, {Neronov}, {Scaife}  \& {Sokolenko}}{{Ar{\'a}mburo-Garc{\'\i}a}
  et~al.}{2023}]{aramburo2023}
{Ar{\'a}mburo-Garc{\'\i}a} A.,  {Bondarenko} K.,  {Boyarsky} A.,  {Neronov} A.,
   {Scaife} A.,   {Sokolenko} A.,  2023, \mn@doi [\mnras]
  {10.1093/mnras/stac3728}, \href
  {https://ui.adsabs.harvard.edu/abs/2023MNRAS.519.4030A} {519, 4030}

\bibitem[\protect\citeauthoryear{{Basu}, {Mao}, {Fletcher}, {Kanekar},
  {Shukurov}, {Schnitzeler}, {Vacca}  \& {Junklewitz}}{{Basu}
  et~al.}{2018}]{basu2018}
{Basu} A.,  {Mao} S.~A.,  {Fletcher} A.,  {Kanekar} N.,  {Shukurov} A.,
  {Schnitzeler} D.,  {Vacca} V.,   {Junklewitz} H.,  2018, \mn@doi [\mnras]
  {10.1093/mnras/sty766}, \href
  {https://ui.adsabs.harvard.edu/abs/2018MNRAS.477.2528B} {477, 2528}

\bibitem[\protect\citeauthoryear{{Berlok} \& {Pfrommer}}{{Berlok} \&
  {Pfrommer}}{2019}]{berlok2019}
{Berlok} T.,  {Pfrommer} C.,  2019, \mn@doi [\mnras] {10.1093/mnras/stz2347},
  \href {https://ui.adsabs.harvard.edu/abs/2019MNRAS.489.3368B} {489, 3368}

\bibitem[\protect\citeauthoryear{{Butsky} et~al.,}{{Butsky}
  et~al.}{2022}]{butsky2022}
{Butsky} I.~S.,  et~al., 2022, \mn@doi [\apj] {10.3847/1538-4357/ac7ebd}, \href
  {https://ui.adsabs.harvard.edu/abs/2022ApJ...935...69B} {935, 69}

\bibitem[\protect\citeauthoryear{{Cameron} et~al.,}{{Cameron}
  et~al.}{2021}]{cameron2021}
{Cameron} A.~J.,  et~al., 2021, \mn@doi [\apjl] {10.3847/2041-8213/ac18ca},
  \href {https://ui.adsabs.harvard.edu/abs/2021ApJ...918L..16C} {918, L16}

\bibitem[\protect\citeauthoryear{{Carretti}, {O'Sullivan}, {Vacca}, {Vazza},
  {Gheller}, {Vernstrom}  \& {Bonafede}}{{Carretti}
  et~al.}{2023}]{carretti2023}
{Carretti} E.,  {O'Sullivan} S.~P.,  {Vacca} V.,  {Vazza} F.,  {Gheller} C.,
  {Vernstrom} T.,   {Bonafede} A.,  2023, \mn@doi [\mnras]
  {10.1093/mnras/stac2966}, \href
  {https://ui.adsabs.harvard.edu/abs/2023MNRAS.518.2273C} {518, 2273}

\bibitem[\protect\citeauthoryear{{Chadayammuri}, {Bogd{\'a}n}, {Oppenheimer},
  {Kraft}, {Forman}  \& {Jones}}{{Chadayammuri}
  et~al.}{2022}]{chadayammuri2022}
{Chadayammuri} U.,  {Bogd{\'a}n} {\'A}.,  {Oppenheimer} B.~D.,  {Kraft} R.~P.,
  {Forman} W.~R.,   {Jones} C.,  2022, \mn@doi [\apjl]
  {10.3847/2041-8213/ac8936}, \href
  {https://ui.adsabs.harvard.edu/abs/2022ApJ...936L..15C} {936, L15}

\bibitem[\protect\citeauthoryear{{Comparat} et~al.,}{{Comparat}
  et~al.}{2022}]{comparat2022}
{Comparat} J.,  et~al., 2022, \mn@doi [\aap] {10.1051/0004-6361/202243101},
  \href {https://ui.adsabs.harvard.edu/abs/2022A&A...666A.156C} {666, A156}

\bibitem[\protect\citeauthoryear{{Davies}, {Crain}, {Oppenheimer}  \&
  {Schaye}}{{Davies} et~al.}{2020}]{davies2020}
{Davies} J.~J.,  {Crain} R.~A.,  {Oppenheimer} B.~D.,   {Schaye} J.,  2020,
  \mn@doi [\mnras] {10.1093/mnras/stz3201}, \href
  {https://ui.adsabs.harvard.edu/abs/2020MNRAS.491.4462D} {491, 4462}

\bibitem[\protect\citeauthoryear{{Dolag}, {Grasso}, {Springel}  \&
  {Tkachev}}{{Dolag} et~al.}{2005}]{dolag2005}
{Dolag} K.,  {Grasso} D.,  {Springel} V.,   {Tkachev} I.,  2005, \mn@doi
  [\jcap] {10.1088/1475-7516/2005/01/009}, \href
  {https://ui.adsabs.harvard.edu/abs/2005JCAP...01..009D} {2005, 009}

\bibitem[\protect\citeauthoryear{{Donahue} \& {Voit}}{{Donahue} \&
  {Voit}}{2022}]{donahue2022}
{Donahue} M.,  {Voit} G.~M.,  2022, \mn@doi [\physrep]
  {10.1016/j.physrep.2022.04.005}, \href
  {https://ui.adsabs.harvard.edu/abs/2022PhR...973....1D} {973, 1}

\bibitem[\protect\citeauthoryear{{Fielding}, {Quataert}, {McCourt}  \&
  {Thompson}}{{Fielding} et~al.}{2017}]{fielding2017}
{Fielding} D.,  {Quataert} E.,  {McCourt} M.,   {Thompson} T.~A.,  2017,
  \mn@doi [\mnras] {10.1093/mnras/stw3326}, \href
  {https://ui.adsabs.harvard.edu/abs/2017MNRAS.466.3810F} {466, 3810}

\bibitem[\protect\citeauthoryear{{Gaensler}, {Landecker}, {Taylor}  \& {POSSUM
  Collaboration}}{{Gaensler} et~al.}{2010}]{gaensler2010}
{Gaensler} B.~M.,  {Landecker} T.~L.,  {Taylor} A.~R.,   {POSSUM Collaboration}
  2010, in American Astronomical Society Meeting Abstracts \#215. p. 470.13

\bibitem[\protect\citeauthoryear{{Garaldi}, {Pakmor}  \& {Springel}}{{Garaldi}
  et~al.}{2021}]{garaldi2021}
{Garaldi} E.,  {Pakmor} R.,   {Springel} V.,  2021, \mn@doi [\mnras]
  {10.1093/mnras/stab086}, \href
  {https://ui.adsabs.harvard.edu/abs/2021MNRAS.502.5726G} {502, 5726}

\bibitem[\protect\citeauthoryear{{Hackstein}, {Br{\"u}ggen}, {Vazza},
  {Gaensler}  \& {Heesen}}{{Hackstein} et~al.}{2019}]{hackstein2019}
{Hackstein} S.,  {Br{\"u}ggen} M.,  {Vazza} F.,  {Gaensler} B.~M.,   {Heesen}
  V.,  2019, \mn@doi [\mnras] {10.1093/mnras/stz2033}, \href
  {https://ui.adsabs.harvard.edu/abs/2019MNRAS.488.4220H} {488, 4220}

\bibitem[\protect\citeauthoryear{{Hackstein}, {Br{\"u}ggen}, {Vazza}  \&
  {Rodrigues}}{{Hackstein} et~al.}{2020}]{hackstein2020}
{Hackstein} S.,  {Br{\"u}ggen} M.,  {Vazza} F.,   {Rodrigues} L.~F.~S.,  2020,
  \mn@doi [\mnras] {10.1093/mnras/staa2572}, \href
  {https://ui.adsabs.harvard.edu/abs/2020MNRAS.498.4811H} {498, 4811}

\bibitem[\protect\citeauthoryear{{Heald} et~al.,}{{Heald}
  et~al.}{2020}]{heald2020}
{Heald} G.,  et~al., 2020, \mn@doi [Galaxies] {10.3390/galaxies8030053}, \href
  {https://ui.adsabs.harvard.edu/abs/2020Galax...8...53H} {8, 53}

\bibitem[\protect\citeauthoryear{{Heesen} et~al.,}{{Heesen}
  et~al.}{2023}]{heesen2023}
{Heesen} V.,  et~al., 2023, \mn@doi [\aap] {10.1051/0004-6361/202346008}, \href
  {https://ui.adsabs.harvard.edu/abs/2023A&A...670L..23H} {670, L23}

\bibitem[\protect\citeauthoryear{{Hilmarsson} et~al.,}{{Hilmarsson}
  et~al.}{2021}]{hilmarsson2021}
{Hilmarsson} G.~H.,  et~al., 2021, \mn@doi [\apjl] {10.3847/2041-8213/abdec0},
  \href {https://ui.adsabs.harvard.edu/abs/2021ApJ...908L..10H} {908, L10}

\bibitem[\protect\citeauthoryear{{Hopkins} \& {Raives}}{{Hopkins} \&
  {Raives}}{2016}]{hopkins2016}
{Hopkins} P.~F.,  {Raives} M.~J.,  2016, \mn@doi [\mnras]
  {10.1093/mnras/stv2180}, \href
  {https://ui.adsabs.harvard.edu/abs/2016MNRAS.455...51H} {455, 51}

\bibitem[\protect\citeauthoryear{{Ideguchi}, {Tashiro}, {Akahori}, {Takahashi}
  \& {Ryu}}{{Ideguchi} et~al.}{2017}]{ideguchi2017}
{Ideguchi} S.,  {Tashiro} Y.,  {Akahori} T.,  {Takahashi} K.,   {Ryu} D.,
  2017, \mn@doi [\apj] {10.3847/1538-4357/aa79a1}, \href
  {https://ui.adsabs.harvard.edu/abs/2017ApJ...843..146I} {843, 146}

\bibitem[\protect\citeauthoryear{{Ji}, {Oh}  \& {McCourt}}{{Ji}
  et~al.}{2018}]{ji2018}
{Ji} S.,  {Oh} S.~P.,   {McCourt} M.,  2018, \mn@doi [\mnras]
  {10.1093/mnras/sty293}, \href
  {https://ui.adsabs.harvard.edu/abs/2018MNRAS.476..852J} {476, 852}

\bibitem[\protect\citeauthoryear{{Kim}, {Lilly}, {Miniati}, {Bernet}, {Beck},
  {O'Sullivan}  \& {Gaensler}}{{Kim} et~al.}{2016}]{kim2016}
{Kim} K.~S.,  {Lilly} S.~J.,  {Miniati} F.,  {Bernet} M.~L.,  {Beck} R.,
  {O'Sullivan} S.~P.,   {Gaensler} B.~M.,  2016, \mn@doi [\apj]
  {10.3847/0004-637X/829/2/133}, \href
  {https://ui.adsabs.harvard.edu/abs/2016ApJ...829..133K} {829, 133}

\bibitem[\protect\citeauthoryear{{Lan} \& {Prochaska}}{{Lan} \&
  {Prochaska}}{2020}]{lan2020}
{Lan} T.-W.,  {Prochaska} J.~X.,  2020, \mn@doi [\mnras]
  {10.1093/mnras/staa1750}, \href
  {https://ui.adsabs.harvard.edu/abs/2020MNRAS.496.3142L} {496, 3142}

\bibitem[\protect\citeauthoryear{{Lerche}}{{Lerche}}{1970}]{lerche1970}
{Lerche} I.,  1970, \mn@doi [\apss] {10.1007/BF00653866}, \href
  {https://ui.adsabs.harvard.edu/abs/1970Ap&SS...6..481L} {6, 481}

\bibitem[\protect\citeauthoryear{{Li} \& {Tonnesen}}{{Li} \&
  {Tonnesen}}{2020}]{li2020}
{Li} M.,  {Tonnesen} S.,  2020, \mn@doi [\apj] {10.3847/1538-4357/ab9f9f},
  \href {https://ui.adsabs.harvard.edu/abs/2020ApJ...898..148L} {898, 148}

\bibitem[\protect\citeauthoryear{{Liu}, {Kretschmer}  \& {Teyssier}}{{Liu}
  et~al.}{2022}]{liu2022}
{Liu} Y.,  {Kretschmer} M.,   {Teyssier} R.,  2022, \mn@doi [\mnras]
  {10.1093/mnras/stac1266}, \href
  {https://ui.adsabs.harvard.edu/abs/2022MNRAS.513.6028L} {513, 6028}

\bibitem[\protect\citeauthoryear{{Mannings} et~al.,}{{Mannings}
  et~al.}{2022}]{mannings2022}
{Mannings} A.~G.,  et~al., 2022, \mn@doi [arXiv e-prints]
  {10.48550/arXiv.2209.15113}, \href
  {https://ui.adsabs.harvard.edu/abs/2022arXiv220915113M} {p. arXiv:2209.15113}

\bibitem[\protect\citeauthoryear{{Mao} et~al.,}{{Mao} et~al.}{2017}]{mao2017}
{Mao} S.~A.,  et~al., 2017, \mn@doi [Nature Astronomy]
  {10.1038/s41550-017-0218-x}, \href
  {https://ui.adsabs.harvard.edu/abs/2017NatAs...1..621M} {1, 621}

\bibitem[\protect\citeauthoryear{{Marinacci}, {Vogelsberger}, {Mocz}  \&
  {Pakmor}}{{Marinacci} et~al.}{2015}]{marinacci2015}
{Marinacci} F.,  {Vogelsberger} M.,  {Mocz} P.,   {Pakmor} R.,  2015, \mn@doi
  [\mnras] {10.1093/mnras/stv1692}, \href
  {https://ui.adsabs.harvard.edu/abs/2015MNRAS.453.3999M} {453, 3999}

\bibitem[\protect\citeauthoryear{Marinacci et~al.,}{Marinacci
  et~al.}{2018}]{marinacci2018}
Marinacci F.,  et~al., 2018, Monthly Notices of the Royal Astronomical Society,
  480, 5113

\bibitem[\protect\citeauthoryear{{Martin-Alvarez}, {Katz}, {Sijacki},
  {Devriendt}  \& {Slyz}}{{Martin-Alvarez} et~al.}{2021}]{martinalvarez2021}
{Martin-Alvarez} S.,  {Katz} H.,  {Sijacki} D.,  {Devriendt} J.,   {Slyz} A.,
  2021, \mn@doi [\mnras] {10.1093/mnras/stab968}, \href
  {https://ui.adsabs.harvard.edu/abs/2021MNRAS.504.2517M} {504, 2517}

\bibitem[\protect\citeauthoryear{{Mart{\'\i}n-Navarro}, {Pillepich}, {Nelson},
  {Rodriguez-Gomez}, {Donnari}, {Hernquist}  \&
  {Springel}}{{Mart{\'\i}n-Navarro} et~al.}{2021}]{martinnavarro2021}
{Mart{\'\i}n-Navarro} I.,  {Pillepich} A.,  {Nelson} D.,  {Rodriguez-Gomez} V.,
   {Donnari} M.,  {Hernquist} L.,   {Springel} V.,  2021, \mn@doi [\nat]
  {10.1038/s41586-021-03545-9}, \href
  {https://ui.adsabs.harvard.edu/abs/2021Natur.594..187M} {594, 187}

\bibitem[\protect\citeauthoryear{{Mitchell}, {Schaye}, {Bower}  \&
  {Crain}}{{Mitchell} et~al.}{2020}]{mitchell2020}
{Mitchell} P.~D.,  {Schaye} J.,  {Bower} R.~G.,   {Crain} R.~A.,  2020, \mn@doi
  [\mnras] {10.1093/mnras/staa938}, \href
  {https://ui.adsabs.harvard.edu/abs/2020MNRAS.494.3971M} {494, 3971}

\bibitem[\protect\citeauthoryear{Naiman et~al.,}{Naiman
  et~al.}{2018}]{naiman2018}
Naiman J.~P.,  et~al., 2018, Monthly Notices of the Royal Astronomical Society,
  477, 1206

\bibitem[\protect\citeauthoryear{Nelson et~al.,}{Nelson
  et~al.}{2018a}]{nelson2018}
Nelson D.,  et~al., 2018a, Monthly Notices of the Royal Astronomical Society,
  475, 624

\bibitem[\protect\citeauthoryear{{Nelson} et~al.,}{{Nelson}
  et~al.}{2018b}]{nelson2018b}
{Nelson} D.,  et~al., 2018b, \mn@doi [\mnras] {10.1093/mnras/sty656}, \href
  {https://ui.adsabs.harvard.edu/abs/2018MNRAS.477..450N} {477, 450}

\bibitem[\protect\citeauthoryear{{Nelson} et~al.,}{{Nelson}
  et~al.}{2019a}]{nelson2019b}
{Nelson} D.,  et~al., 2019a, \mn@doi [Computational Astrophysics and Cosmology]
  {10.1186/s40668-019-0028-x}, \href
  {https://ui.adsabs.harvard.edu/abs/2019ComAC...6....2N} {6, 2}

\bibitem[\protect\citeauthoryear{{Nelson} et~al.,}{{Nelson}
  et~al.}{2019b}]{nelson2019}
{Nelson} D.,  et~al., 2019b, \mn@doi [\mnras] {10.1093/mnras/stz2306}, \href
  {https://ui.adsabs.harvard.edu/abs/2019MNRAS.490.3234N} {490, 3234}

\bibitem[\protect\citeauthoryear{{Nelson} et~al.,}{{Nelson}
  et~al.}{2020}]{nelson2020}
{Nelson} D.,  et~al., 2020, \mn@doi [\mnras] {10.1093/mnras/staa2419}, \href
  {https://ui.adsabs.harvard.edu/abs/2020MNRAS.498.2391N} {498, 2391}

\bibitem[\protect\citeauthoryear{{O'Sullivan} et~al.,}{{O'Sullivan}
  et~al.}{2020}]{osullivan2020}
{O'Sullivan} S.~P.,  et~al., 2020, \mn@doi [\mnras] {10.1093/mnras/staa1395},
  \href {https://ui.adsabs.harvard.edu/abs/2020MNRAS.495.2607O} {495, 2607}

\bibitem[\protect\citeauthoryear{{O'Sullivan} et~al.,}{{O'Sullivan}
  et~al.}{2023}]{osullivan2023}
{O'Sullivan} S.~P.,  et~al., 2023, \mn@doi [\mnras] {10.1093/mnras/stac3820},
  \href {https://ui.adsabs.harvard.edu/abs/2023MNRAS.519.5723O} {519, 5723}

\bibitem[\protect\citeauthoryear{{Oppenheimer} et~al.,}{{Oppenheimer}
  et~al.}{2020}]{oppenheimer2020}
{Oppenheimer} B.~D.,  et~al., 2020, \mn@doi [\mnras] {10.1093/mnras/stz3124},
  \href {https://ui.adsabs.harvard.edu/abs/2020MNRAS.491.2939O} {491, 2939}

\bibitem[\protect\citeauthoryear{{Pakmor} \& {Springel}}{{Pakmor} \&
  {Springel}}{2013}]{pakmor2013}
{Pakmor} R.,  {Springel} V.,  2013, \mn@doi [\mnras] {10.1093/mnras/stt428},
  \href {https://ui.adsabs.harvard.edu/abs/2013MNRAS.432..176P} {432, 176}

\bibitem[\protect\citeauthoryear{{Pakmor}, {Bauer}  \& {Springel}}{{Pakmor}
  et~al.}{2011}]{pakmor2011}
{Pakmor} R.,  {Bauer} A.,   {Springel} V.,  2011, \mn@doi [\mnras]
  {10.1111/j.1365-2966.2011.19591.x}, \href
  {https://ui.adsabs.harvard.edu/abs/2011MNRAS.418.1392P} {418, 1392}

\bibitem[\protect\citeauthoryear{{Pakmor}, {Marinacci}  \& {Springel}}{{Pakmor}
  et~al.}{2014}]{pakmor2014}
{Pakmor} R.,  {Marinacci} F.,   {Springel} V.,  2014, \mn@doi [\apjl]
  {10.1088/2041-8205/783/1/L20}, \href
  {https://ui.adsabs.harvard.edu/abs/2014ApJ...783L..20P} {783, L20}

\bibitem[\protect\citeauthoryear{{Pakmor} et~al.,}{{Pakmor}
  et~al.}{2017}]{pakmor2017}
{Pakmor} R.,  et~al., 2017, \mn@doi [\mnras] {10.1093/mnras/stx1074}, \href
  {https://ui.adsabs.harvard.edu/abs/2017MNRAS.469.3185P} {469, 3185}

\bibitem[\protect\citeauthoryear{{Pakmor} et~al.,}{{Pakmor}
  et~al.}{2020}]{pakmor2020}
{Pakmor} R.,  et~al., 2020, \mn@doi [\mnras] {10.1093/mnras/staa2530}, \href
  {https://ui.adsabs.harvard.edu/abs/2020MNRAS.498.3125P} {498, 3125}

\bibitem[\protect\citeauthoryear{{Pandya} et~al.,}{{Pandya}
  et~al.}{2021}]{pandya2021}
{Pandya} V.,  et~al., 2021, \mn@doi [\mnras] {10.1093/mnras/stab2714}, \href
  {https://ui.adsabs.harvard.edu/abs/2021MNRAS.508.2979P} {508, 2979}

\bibitem[\protect\citeauthoryear{{P{\'e}roux}, {Nelson}, {van de Voort},
  {Pillepich}, {Marinacci}, {Vogelsberger}  \& {Hernquist}}{{P{\'e}roux}
  et~al.}{2020}]{peroux2020}
{P{\'e}roux} C.,  {Nelson} D.,  {van de Voort} F.,  {Pillepich} A.,
  {Marinacci} F.,  {Vogelsberger} M.,   {Hernquist} L.,  2020, \mn@doi [\mnras]
  {10.1093/mnras/staa2888}, \href
  {https://ui.adsabs.harvard.edu/abs/2020MNRAS.499.2462P} {499, 2462}

\bibitem[\protect\citeauthoryear{Pillepich et~al.,}{Pillepich
  et~al.}{2018a}]{pillepich2018a}
Pillepich A.,  et~al., 2018a, Monthly Notices of the Royal Astronomical
  Society, 473, 4077

\bibitem[\protect\citeauthoryear{Pillepich et~al.,}{Pillepich
  et~al.}{2018b}]{pillepich2018b}
Pillepich A.,  et~al., 2018b, Monthly Notices of the Royal Astronomical
  Society, 475, 648

\bibitem[\protect\citeauthoryear{{Pillepich} et~al.,}{{Pillepich}
  et~al.}{2019}]{pillepich2019}
{Pillepich} A.,  et~al., 2019, \mn@doi [\mnras] {10.1093/mnras/stz2338}, \href
  {https://ui.adsabs.harvard.edu/abs/2019MNRAS.490.3196P} {490, 3196}

\bibitem[\protect\citeauthoryear{{Pillepich}, {Nelson}, {Truong}, {Weinberger},
  {Martin-Navarro}, {Springel}, {Faber}  \& {Hernquist}}{{Pillepich}
  et~al.}{2021}]{pillepich2021}
{Pillepich} A.,  {Nelson} D.,  {Truong} N.,  {Weinberger} R.,  {Martin-Navarro}
  I.,  {Springel} V.,  {Faber} S.~M.,   {Hernquist} L.,  2021, \mn@doi [\mnras]
  {10.1093/mnras/stab2779}, \href
  {https://ui.adsabs.harvard.edu/abs/2021MNRAS.508.4667P} {508, 4667}

\bibitem[\protect\citeauthoryear{{Planck Collaboration} et~al.,}{{Planck
  Collaboration} et~al.}{2016}]{planck2016}
{Planck Collaboration} et~al., 2016, Astronomy \& Astrophysics, 594, A13

\bibitem[\protect\citeauthoryear{{Pomakov} et~al.,}{{Pomakov}
  et~al.}{2022}]{2022MNRAS.515..256P}
{Pomakov} V.~P.,  et~al., 2022, \mn@doi [\mnras] {10.1093/mnras/stac1805},
  \href {https://ui.adsabs.harvard.edu/abs/2022MNRAS.515..256P} {515, 256}

\bibitem[\protect\citeauthoryear{{Ponnada} et~al.,}{{Ponnada}
  et~al.}{2022}]{ponnada2022}
{Ponnada} S.~B.,  et~al., 2022, \mn@doi [\mnras] {10.1093/mnras/stac2448},
  \href {https://ui.adsabs.harvard.edu/abs/2022MNRAS.516.4417P} {516, 4417}

\bibitem[\protect\citeauthoryear{{Powell}, {Roe}, {Linde}, {Gombosi}  \& {De
  Zeeuw}}{{Powell} et~al.}{1999}]{powell1999}
{Powell} K.~G.,  {Roe} P.~L.,  {Linde} T.~J.,  {Gombosi} T.~I.,   {De Zeeuw}
  D.~L.,  1999, \mn@doi [Journal of Computational Physics]
  {10.1006/jcph.1999.6299}, \href
  {https://ui.adsabs.harvard.edu/abs/1999JCoPh.154..284P} {154, 284}

\bibitem[\protect\citeauthoryear{Predehl et~al.,}{Predehl
  et~al.}{2020}]{predehl2020}
Predehl P.,  et~al., 2020, \mn@doi [Nature] {10.1038/s41586-020-2979-0}, 588,
  227

\bibitem[\protect\citeauthoryear{{Prochaska} et~al.,}{{Prochaska}
  et~al.}{2019}]{prochaska2019}
{Prochaska} J.~X.,  et~al., 2019, \mn@doi [Science] {10.1126/science.aay0073},
  \href {https://ui.adsabs.harvard.edu/abs/2019Sci...366..231P} {366, 231}

\bibitem[\protect\citeauthoryear{{Ramesh} \& {Nelson}}{{Ramesh} \&
  {Nelson}}{2023}]{ramesh2023d}
{Ramesh} R.,  {Nelson} D.,  2023, \mn@doi [arXiv e-prints]
  {10.48550/arXiv.2307.11143}, \href
  {https://ui.adsabs.harvard.edu/abs/2023arXiv230711143R} {p. arXiv:2307.11143}

\bibitem[\protect\citeauthoryear{{Ramesh}, {Nelson}  \& {Pillepich}}{{Ramesh}
  et~al.}{2023a}]{ramesh2023a}
{Ramesh} R.,  {Nelson} D.,   {Pillepich} A.,  2023a, \mn@doi [\mnras]
  {10.1093/mnras/stac3524}, \href
  {https://ui.adsabs.harvard.edu/abs/2023MNRAS.518.5754R} {518, 5754}

\bibitem[\protect\citeauthoryear{{Ramesh}, {Nelson}  \& {Pillepich}}{{Ramesh}
  et~al.}{2023b}]{ramesh2023b}
{Ramesh} R.,  {Nelson} D.,   {Pillepich} A.,  2023b, \mn@doi [\mnras]
  {10.1093/mnras/stad951}, \href
  {https://ui.adsabs.harvard.edu/abs/2023MNRAS.522.1535R} {522, 1535}

\bibitem[\protect\citeauthoryear{{Rodriguez-Gomez} et~al.,}{{Rodriguez-Gomez}
  et~al.}{2019}]{rodriguezgomez2019}
{Rodriguez-Gomez} V.,  et~al., 2019, \mn@doi [\mnras] {10.1093/mnras/sty3345},
  \href {https://ui.adsabs.harvard.edu/abs/2019MNRAS.483.4140R} {483, 4140}

\bibitem[\protect\citeauthoryear{{Rudnick} \& {Cotton}}{{Rudnick} \&
  {Cotton}}{2023}]{rudnick2023}
{Rudnick} L.,  {Cotton} W.~D.,  2023, \mn@doi [arXiv e-prints]
  {10.48550/arXiv.2304.02728}, \href
  {https://ui.adsabs.harvard.edu/abs/2023arXiv230402728R} {p. arXiv:2304.02728}

\bibitem[\protect\citeauthoryear{{Seta} \& {Federrath}}{{Seta} \&
  {Federrath}}{2021}]{seta2021}
{Seta} A.,  {Federrath} C.,  2021, \mn@doi [\mnras] {10.1093/mnras/stab128},
  \href {https://ui.adsabs.harvard.edu/abs/2021MNRAS.502.2220S} {502, 2220}

\bibitem[\protect\citeauthoryear{{Shah} \& {Seta}}{{Shah} \&
  {Seta}}{2021}]{shah2021}
{Shah} H.,  {Seta} A.,  2021, \mn@doi [\mnras] {10.1093/mnras/stab2500}, \href
  {https://ui.adsabs.harvard.edu/abs/2021MNRAS.508.1371S} {508, 1371}

\bibitem[\protect\citeauthoryear{{Sobey} et~al.,}{{Sobey}
  et~al.}{2019}]{sobey2019}
{Sobey} C.,  et~al., 2019, \mn@doi [\mnras] {10.1093/mnras/stz214}, \href
  {https://ui.adsabs.harvard.edu/abs/2019MNRAS.484.3646S} {484, 3646}

\bibitem[\protect\citeauthoryear{{Sparre}, {Pfrommer}  \& {Ehlert}}{{Sparre}
  et~al.}{2020}]{sparre2020}
{Sparre} M.,  {Pfrommer} C.,   {Ehlert} K.,  2020, \mn@doi [\mnras]
  {10.1093/mnras/staa3177}, \href
  {https://ui.adsabs.harvard.edu/abs/2020MNRAS.499.4261S} {499, 4261}

\bibitem[\protect\citeauthoryear{Springel}{Springel}{2010}]{springel2010}
Springel V.,  2010, Monthly Notices of the Royal Astronomical Society, 401, 791

\bibitem[\protect\citeauthoryear{{Springel} \& {Hernquist}}{{Springel} \&
  {Hernquist}}{2003}]{springel2003}
{Springel} V.,  {Hernquist} L.,  2003, \mn@doi [\mnras]
  {10.1046/j.1365-8711.2003.06206.x}, \href
  {https://ui.adsabs.harvard.edu/abs/2003MNRAS.339..289S} {339, 289}

\bibitem[\protect\citeauthoryear{{Springel}, {White}, {Tormen}  \&
  {Kauffmann}}{{Springel} et~al.}{2001}]{springel2001}
{Springel} V.,  {White} S. D.~M.,  {Tormen} G.,   {Kauffmann} G.,  2001,
  \mn@doi [\mnras] {10.1046/j.1365-8711.2001.04912.x}, \href
  {https://ui.adsabs.harvard.edu/abs/2001MNRAS.328..726S} {328, 726}

\bibitem[\protect\citeauthoryear{Springel et~al.,}{Springel
  et~al.}{2018}]{springel2018}
Springel V.,  et~al., 2018, Monthly Notices of the Royal Astronomical Society,
  475, 676

\bibitem[\protect\citeauthoryear{{Stasyszyn} \& {de los Rios}}{{Stasyszyn} \&
  {de los Rios}}{2019}]{stasyszyn2019}
{Stasyszyn} F.~A.,  {de los Rios} M.,  2019, \mn@doi [\mnras]
  {10.1093/mnras/stz1450}, \href
  {https://ui.adsabs.harvard.edu/abs/2019MNRAS.487.4768S} {487, 4768}

\bibitem[\protect\citeauthoryear{{Steinwandel}, {Dolag}, {Lesch}, {Moster},
  {Burkert}  \& {Prieto}}{{Steinwandel} et~al.}{2020}]{steinwandel2020}
{Steinwandel} U.~P.,  {Dolag} K.,  {Lesch} H.,  {Moster} B.~P.,  {Burkert} A.,
   {Prieto} A.,  2020, \mn@doi [\mnras] {10.1093/mnras/staa817}, \href
  {https://ui.adsabs.harvard.edu/abs/2020MNRAS.494.4393S} {494, 4393}

\bibitem[\protect\citeauthoryear{Su, Slatyer  \& Finkbeiner}{Su
  et~al.}{2010}]{su2010}
Su M.,  Slatyer T.~R.,   Finkbeiner D.~P.,  2010, \mn@doi [ApJ]
  {10.1088/0004-637x/724/2/1044}, 724, 1044

\bibitem[\protect\citeauthoryear{{Takahashi}}{{Takahashi}}{2023}]{takahashi2023}
{Takahashi} K.,  2023, \mn@doi [\pasj] {10.1093/pasj/psac111}, \href
  {https://ui.adsabs.harvard.edu/abs/2023PASJ...75S..50T} {75, S50}

\bibitem[\protect\citeauthoryear{{Truong} et~al.,}{{Truong}
  et~al.}{2020}]{truong2020}
{Truong} N.,  et~al., 2020, \mn@doi [\mnras] {10.1093/mnras/staa685}, \href
  {https://ui.adsabs.harvard.edu/abs/2020MNRAS.494..549T} {494, 549}

\bibitem[\protect\citeauthoryear{{Truong}, {Pillepich}, {Nelson}, {Werner}  \&
  {Hernquist}}{{Truong} et~al.}{2021}]{truong2021}
{Truong} N.,  {Pillepich} A.,  {Nelson} D.,  {Werner} N.,   {Hernquist} L.,
  2021, \mn@doi [\mnras] {10.1093/mnras/stab2638}, \href
  {https://ui.adsabs.harvard.edu/abs/2021MNRAS.508.1563T} {508, 1563}

\bibitem[\protect\citeauthoryear{{Vacca} et~al.,}{{Vacca}
  et~al.}{2016}]{vacca2016}
{Vacca} V.,  et~al., 2016, \mn@doi [\aap] {10.1051/0004-6361/201527291}, \href
  {https://ui.adsabs.harvard.edu/abs/2016A&A...591A..13V} {591, A13}

\bibitem[\protect\citeauthoryear{{Vazza}, {Br{\"u}ggen}, {Gheller}  \&
  {Wang}}{{Vazza} et~al.}{2014}]{vazza2014}
{Vazza} F.,  {Br{\"u}ggen} M.,  {Gheller} C.,   {Wang} P.,  2014, \mn@doi
  [\mnras] {10.1093/mnras/stu1896}, \href
  {https://ui.adsabs.harvard.edu/abs/2014MNRAS.445.3706V} {445, 3706}

\bibitem[\protect\citeauthoryear{{Vazza}, {Br{\"u}ggen}, {Gheller},
  {Hackstein}, {Wittor}  \& {Hinz}}{{Vazza} et~al.}{2017}]{vazza2017}
{Vazza} F.,  {Br{\"u}ggen} M.,  {Gheller} C.,  {Hackstein} S.,  {Wittor} D.,
  {Hinz} P.~M.,  2017, \mn@doi [Classical and Quantum Gravity]
  {10.1088/1361-6382/aa8e60}, \href
  {https://ui.adsabs.harvard.edu/abs/2017CQGra..34w4001V} {34, 234001}

\bibitem[\protect\citeauthoryear{{Weinberger} et~al.,}{{Weinberger}
  et~al.}{2017}]{weinberger2017}
{Weinberger} R.,  et~al., 2017, \mn@doi [\mnras] {10.1093/mnras/stw2944}, \href
  {https://ui.adsabs.harvard.edu/abs/2017MNRAS.465.3291W} {465, 3291}

\bibitem[\protect\citeauthoryear{{Weinberger} et~al.,}{{Weinberger}
  et~al.}{2018}]{weinberger2018}
{Weinberger} R.,  et~al., 2018, \mn@doi [\mnras] {10.1093/mnras/sty1733}, \href
  {https://ui.adsabs.harvard.edu/abs/2018MNRAS.479.4056W} {479, 4056}

\bibitem[\protect\citeauthoryear{{Yang}, {Dav{\'e}}, {Cui}, {Cai}, {Peacock}
  \& {Sorini}}{{Yang} et~al.}{2023}]{yang2023}
{Yang} T.,  {Dav{\'e}} R.,  {Cui} W.,  {Cai} Y.-C.,  {Peacock} J.~A.,
  {Sorini} D.,  2023, \mn@doi [arXiv e-prints] {10.48550/arXiv.2305.00602},
  \href {https://ui.adsabs.harvard.edu/abs/2023arXiv230500602Y} {p.
  arXiv:2305.00602}

\bibitem[\protect\citeauthoryear{{Zhang} \& {Zaritsky}}{{Zhang} \&
  {Zaritsky}}{2022}]{zhang2022}
{Zhang} H.,  {Zaritsky} D.,  2022, \mn@doi [\apj] {10.3847/1538-4357/ac9c64},
  \href {https://ui.adsabs.harvard.edu/abs/2022ApJ...941...18Z} {941, 18}

\bibitem[\protect\citeauthoryear{{Zhang}, {Yan}, {Li}, {Zhang}  \&
  {Wang}}{{Zhang} et~al.}{2021}]{zhang2021}
{Zhang} Z.~J.,  {Yan} K.,  {Li} C.~M.,  {Zhang} G.~Q.,   {Wang} F.~Y.,  2021,
  \mn@doi [\apj] {10.3847/1538-4357/abceb9}, \href
  {https://ui.adsabs.harvard.edu/abs/2021ApJ...906...49Z} {906, 49}

\bibitem[\protect\citeauthoryear{{Zinger} et~al.,}{{Zinger}
  et~al.}{2020}]{zinger2020}
{Zinger} E.,  et~al., 2020, \mn@doi [\mnras] {10.1093/mnras/staa2607}, \href
  {https://ui.adsabs.harvard.edu/abs/2020MNRAS.499..768Z} {499, 768}

\bibitem[\protect\citeauthoryear{{van de Voort}, {Bieri}, {Pakmor},
  {G{\'o}mez}, {Grand}  \& {Marinacci}}{{van de Voort}
  et~al.}{2021}]{vandevoort2021}
{van de Voort} F.,  {Bieri} R.,  {Pakmor} R.,  {G{\'o}mez} F.~A.,  {Grand} R.
  J.~J.,   {Marinacci} F.,  2021, \mn@doi [\mnras] {10.1093/mnras/staa3938},
  \href {https://ui.adsabs.harvard.edu/abs/2021MNRAS.501.4888V} {501, 4888}

\makeatother
\end{thebibliography}

\appendix

\section{Simulated Electron Number Densities}
\begin{figure}
\centering 
\includegraphics[width=0.45\textwidth]{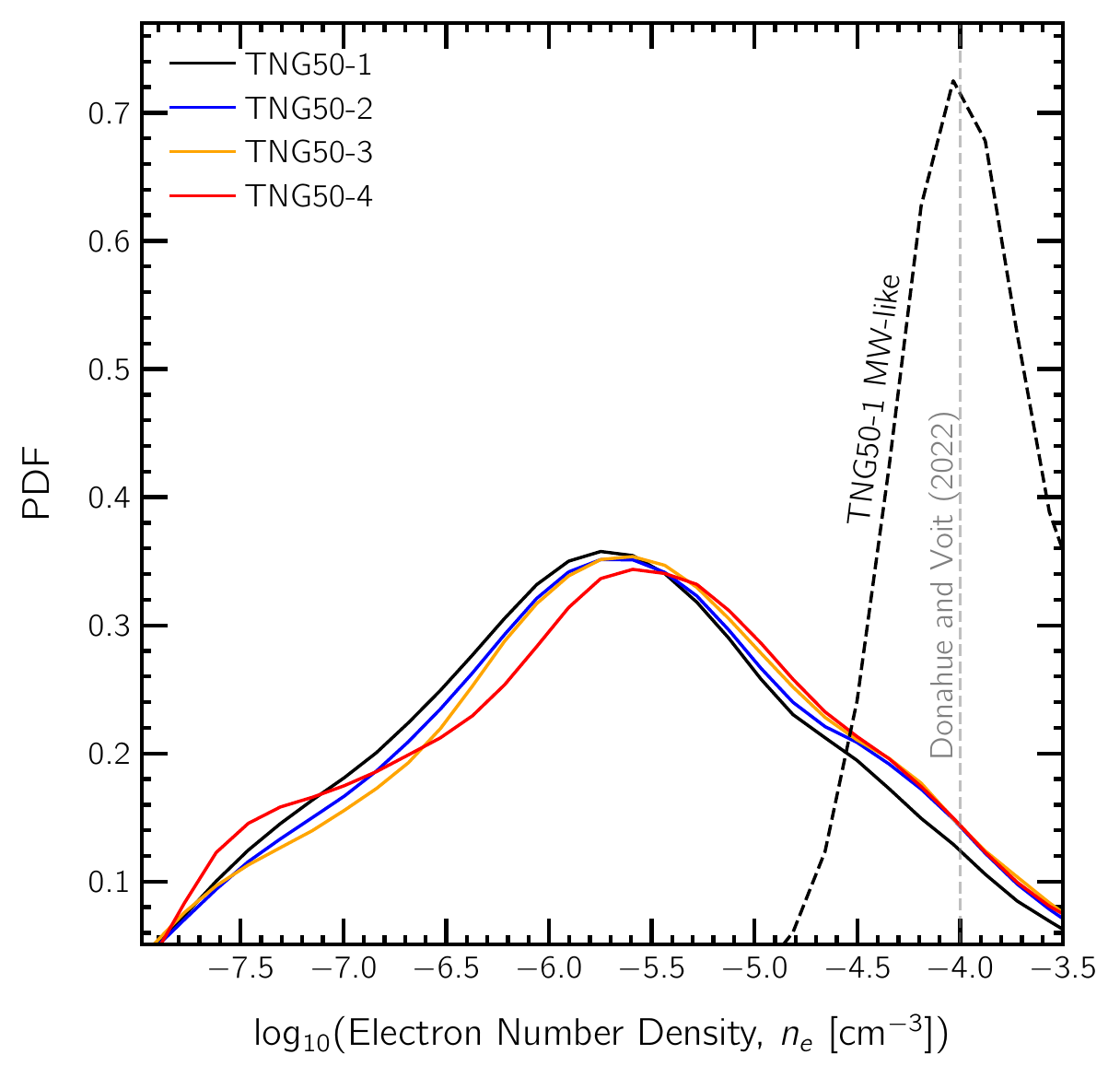}
\caption{PDFs of electron number density of gas cells from the different TNG50 boxes. Except for the low-resolution TNG50-4 box, PDFs show decent convergence. In addition, the dashed vertical gray line shows the best estimate of $n_e$ available from the Milky Way CGM currently \protect\citep{donahue2022}. For a direct comparison, in the dashed black line, we also include the PDF for those gas cells from a sample of TNG50-1 Milky Way-likes, which is in good agreement with the observed value.}
\label{fig:neHist}
\end{figure}

In Figure~\ref{fig:neHist}, we show probability distribution functions (PDFs) of the electron number density of gas cells from the different TNG50 boxes. TNG50-1, which has been used throughout the bulk of this work, is shown in black. The lower resolution boxes of TNG50-2 (mass resolution lower by a factor of 8 with respect to TNG50-1), TNG50-3 (factor of 64) and TNG50-4 (factor of 512) are shown in blue, orange and red, respectively. The peaks of these PDFs shift slightly towards lower values of $n_e$ with improving resolution, although convergence of these distributions between runs is decent barring TNG50-4.

The dashed vertical gray line shows the best observational estimate of $n_e$ currently available from the Milky Way CGM \citep{donahue2022}. This is a value of $10^{-4}$\,cm$^{-3}$, estimated at a galactocentric distance of $\sim$\,$50-100$~kpc. For a direct comparison with this value, in the black dashed line, we show the PDF constructed from a set of Milky Way-like galaxies from TNG50-1 \citep[for details on sample selection, see][]{ramesh2023a}. Similar to the above mentioned observational result, this PDF is restricted to gas cells within a distance of $50-100$~kpc. This PDF peaks close to $\sim$\,$10^{-4}$\,cm$^{-3}$, and is in good agreement with the observed value. Although this comparison is limited to a single observational value, and hence halo mass, this suggests that the under-prediction of RM by TNG is unlikely a result of simulated electron number densities being too low.

\section{Impact of Turbulence and Coherence Lengths}

\begin{figure}
\centering 
\includegraphics[width=0.45\textwidth]{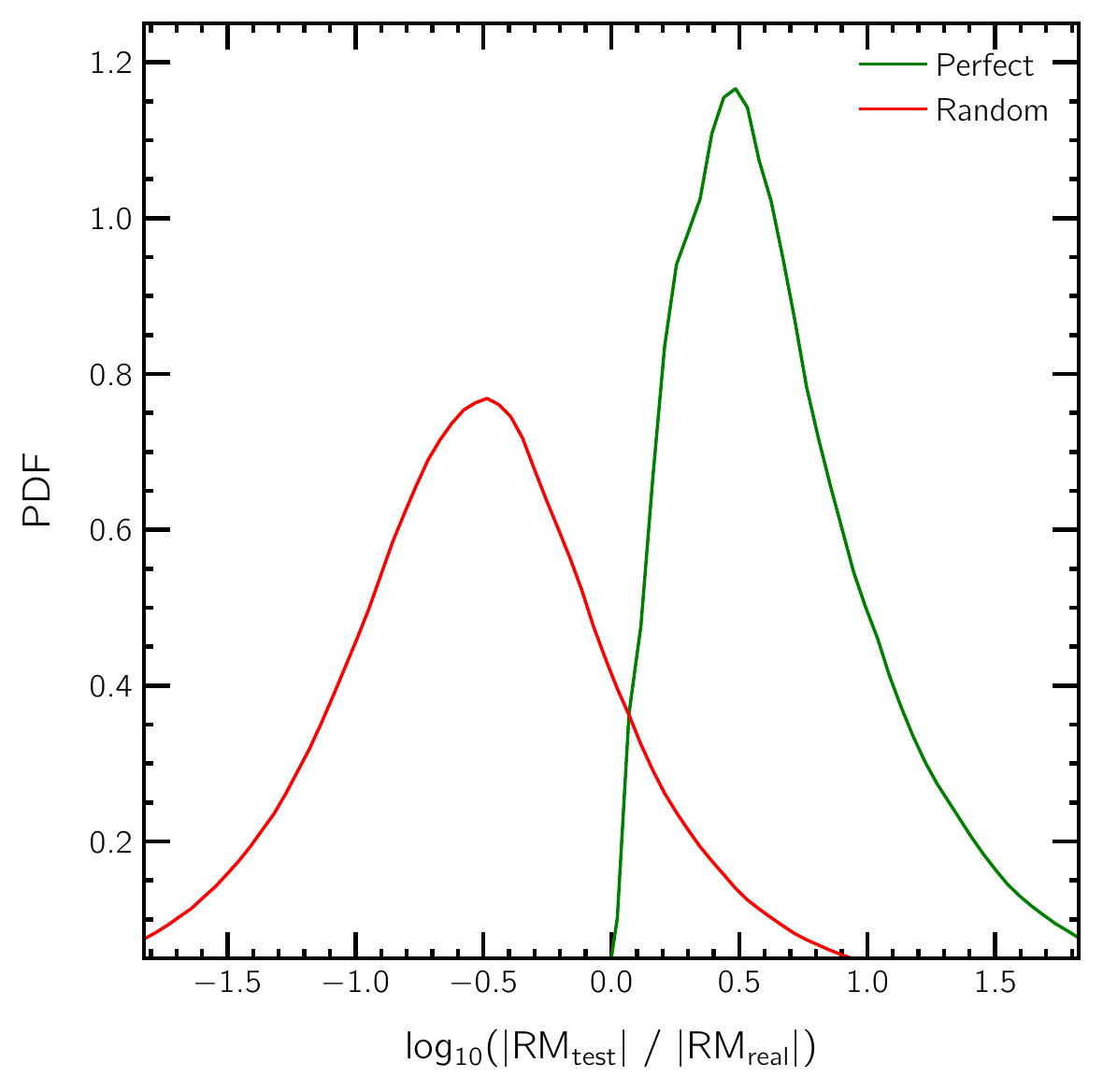}
\caption{Several test cases studying the effect of magnetic coherence length on the obtained RM signal. As explained in the main text, we construct cases where the field lines are perfectly coherent (green) and randomly oriented (red). PDFs of the ratios of these cases with respect to the real case are shown. These tests suggest that the RM signal typically varies by less than an order of magnitude when the coherence length is varied between these two extremes.}
\label{fig:rmTest}
\end{figure}

The measured RM signal is believed to be impacted by factors like the magnetic coherence length \citep[e.g.][]{stasyszyn2019} and turbulence \citep[e.g.][]{lerche1970,ideguchi2017} of gas along the line of sight leading to depolarisation. In Figure~\ref{fig:rmTest}, we explore the impact of such effects on the derived TNG RM values by studying several test cases.

We pick 100 halos at random from $\pm 0.2$~dex bins of $\rm{log_{10}(M_\star)} \in [8.5, 9.0, 9.5, 10.0, 10.5, 11.0, 11.5]$, i.e. a subset of halos from Figure~\ref{fig:b_vs_angle_massBins}. For each of these halos, we calculate three different sets of RM values by varying the magnetic field component along the line of sight ($B_\parallel$):

\begin{enumerate}
    \item Real: using the actual values of $B_\parallel$ from the TNG simulation.
    \item Perfect: assuming perfect alignment of all magnetic field vectors along the line of sight direction, with $B_\parallel$ set to the norm of the magnetic field of the corresponding gas cell.
    \item Random: for each gas cell, we set $B_\parallel$ to the product of the norm of its magnetic field and a random number drawn uniformly from the interval $[-1.0, 1.0]$. This yields a random alignment of magnetic field vectors along the line of sight. 
\end{enumerate}

The perfect alignment case is a proxy for a scenario where the coherence length of B-fields is large, as large as the extent of the halo in this case. The random alignment case probes the other extreme where coherence lengths are small, possibly caused by highly turbulent gas throughout the line of sight interval.\footnote{In our case, small corresponds roughly to the gas spatial resolution in the CGM, i.e. on $\sim$ kpc scales. We cannot assess decoherence on even smaller scales as our method requires a constant magnetic field across a single resolution element.} To assess the impact of these tests specifically on CGM gas, we restrict to those sightlines with impact parameters in the range $[0.15, 1.0] \times \rm{R_{200,c}}$, and a projection column of $\pm \rm{R_{200,c}}$ centred on the galaxy. For each selected sightline, we compute ratios of RM of the test case with respect to the real case. The PDF of these ratios is shown in Figure~\ref{fig:rmTest}: the perfect alignment case in green, and the random alignment in red.

As expected, the green curve shows that a larger coherence length results in a larger RM signal. The typical `strengthening' of the signal is a factor of $\sim$\,$3$, although the values of RM along some sightlines are boosted by factors of $\gtrsim$\,$10$, i.e. by an order of magnitude or more. On the other extreme, randomizing $B_\parallel$ leads to the RM signal growing weaker, typically by a factor of $\sim$\,$0.3$ with respect to the real case. Surprisingly, there are cases where the RM signal grows despite this randomization, suggesting that these sightlines are already dominated by highly turbulent fields.

Overall, while the magnetic coherence length can have an impact on the obtained RM value, the above test cases suggest that the effect is typically less than an order of magnitude. It is thus unlikely that the under-prediction of RM values by TNG50-1 is dominated by this effect.

\end{document}